\documentclass[12pt]{iopart}

\usepackage{subfigure}
\usepackage{graphicx}
\graphicspath{ {./images/} }
\usepackage{threeparttable}
\usepackage{amsfonts}
\usepackage{cite}
\newcommand{\reply}[1]{{{{ #1}}}}

\begin{document}

\title[A spectral method algorithm for numerical simulations of gravitational fields]{A spectral method algorithm for numerical simulations of gravitational fields}

\author{C Meringolo$^1$, S Servidio$^1$ and P Veltri$^1$}

\address{$^1$Dipartimento di Fisica, Universit\`a della Calabria, I-87036 Cosenza, Italy}
\ead{claudiomeringolo@unical.it}
\vspace{10pt}
\begin{indented}
\item[]November 2020
\end{indented}

\begin{abstract}
A numerical study of the Einstein field equations, based on the $3+1$ foliation of the spacetime, is presented. A pseudo-spectral technique has been employed for simulations in vacuum, within two different formalisms, namely the Arnowitt-Deser-Misner (ADM) and the conformal Baumgarte-Shapiro-Shibata-Nakamura (BSSN) approach. The numerical code is based on the Fourier decomposition, accompanied by different filtering techniques. The role of the dealiasing, as well as the influence of the filter type, has been investigated. The algorithms have been stabilized via a novel procedure that controls self-consistently the regularity of the solutions. The accuracy of the model has been validated through standard testbeds, revealing that the filtered pseudo-spectral technique is among the most accurate approaches. Finally, the procedure has been stressed via black hole dynamics and a new strategy, based on hyperviscous dissipation that suppresses spurious boundary problems, has been proposed. The model represents a valid tool of investigation, particularly suitable for the inspection of small scale nonlinear phenomena in gravitational dynamics.

\vspace{15pt}
\end{abstract}

%
%
%
%
%

\section{Introduction}
\label{intro}
In past decades, great effort has been devoted to the numerical study of gravitation. More recently, numerical simulations of the Einstein field equations have been improved significantly. Many of these studies have successfully described the inspiraling of compact objects in their final coalescence phase (and the relative emission of gravitational waves), becoming therefore an extraordinary guidance for the observational campaign of LIGO and VIRGO  \cite{abadie, cutler, bern16, bern94, bern81, Thier, Nakamura, Rezzolla78, abbott1, abbott2}. Several works have been dedicated to the dynamics of Kerr-type black holes, as well as the understanding of the plasma dynamics in their accretion discs. These simulations helped to interpret novel observations, including the celebrated M87 Event Horizon Telescope image of a supermassive black hole \cite{davel, m87, m87_2, m87_3, 14, 9, 13}.

Because of the intrinsic nonlinearity and complexity of the governing equations, several works have been dedicated to the study of both the stability and accuracy of the numerical solutions \cite{Bernuz, Schnetter, campan, poor}. Different forms of the governing equations have been proposed and different numerical strategies have been developed, in a variety of initial conditions known as {\it gravitational testbeds} \cite{Alc}. Many factors affect the dynamics and the accuracy of the solutions, including the formulation, the choice of the gauge conditions, as well as the boundary conditions, the accuracy of the space-time integration, the use of external dissipation and so on \cite{test_1, test_2}.

\reply{
    In this paper, we adopt two different $3+1$ formalisms, based on the standard Arnowitt-Deser-Misner (ADM) decomposition \cite{ADM,admyork} and the conformal Baumgarte-Shapiro-Shibata-Nakamura (BSSN) formulation \cite{50, baumg3, baumg4, DBrown, Sarbach}.
}
We propose a very accurate numerical strategy, in vacuum conditions, based on classical spectral methods. In analogy with high-precision simulations of fluid and plasma dynamics, our numerical code is based upon a Fourier pseudo-spectral method, in concert with filtering techniques. These anti-aliasing techniques are important in direct numerical simulations, since they stabilize the numerical evolution, establishing a direct connection between pseudo-spectral methods and exact Galerkin representations \cite{Servidio_2}. The main advantage of this approach, apart from its simplicity, is the accuracy of the solutions due to the spectral projection, while its limitation relies on the imposition of periodicity at the boundaries. Nevertheless, similar spectral decomposition can be locally used in the flux-reconstruction of finite volume methods, when the 3+1 Einstein field equations are represented in a quasi-hyperbolic form \cite{iper1, iper2}. 

\reply{
    Spectral methods are among the most accurate and performing techniques for numerical relativity. Among all, the SpEC code represents a first, clear example of highly performing spectral algorithm applied to gravitation \cite{ScheelEAL06, Szilagyi14}, where the basic model relies on a first-order, symmetric-hyperbolic, generalized harmonic formulation. Alternatively, in the framework of the BSSN representation, another good example, based on spectral representation of the spherical harmonics, is given by the SGRID code \cite{Tichy09}. Both of the above are multi-domain, pseudo-spectral codes particularly suitable for the study of binary systems. These important, inspiring studies suggest that the spectral decomposition might be both accurate and stable. On this path we proceed here, concentrating on an alternative pseudo-spectral technique and proposing new strategies that prevent numerical pathologies, typical of spectral methods simulations.
}

Apart of the advantage given by the accuracy of the spectral decomposition, the periodic representation can be a valid (and easy) choice for the study of small scale gravitational dynamics, especially in cases where the fields are strongly nonlinear and form local homogeneous patterns, as it can be in the vicinity of the merging-region between compact objects, as well as in the coalescence/collapse of multiple massive bodies \cite{baiotti1,3body, 4body}. Note that, in the case of chaotic patterns with a finite characteristic scale \cite{batch}, periodic conditions can be a valid strategy for the study of small scale dynamics, as commonly done in simulations of turbulent astrophysical plasmas \cite{bill}.

In the present work, the accuracy of our numerical model has been validated through a series of standard tests, like the gauge wave, the robust stability test, the linear gravitational waves and the Gowdy spacetime, as suggested in previous pioneering works \cite{Alc, Dum, bab2, Brown}. The code successfully passed all the numerical tests. Apart from the classical tests, we also proposed a new class of initial data, that satisfies the constraint equations and leads to the formation of standing waves of the metric. In order to minimize the aliasing instability due to nonlinear terms, we propose different anti-aliasing filters, showing how smoother filters might lead to more stable solutions. The algorithms have been stabilized via an adaptive time-refinement, based on a method that self-consistently controls the regularity of the solutions, at each time-step. Finally, we stress the code via black hole dynamics. In order to suppress spurious boundary problems, we propose a new strategy based on hyper-viscous dissipation. 

The work is organized as follows. In section \ref{spacetim} we discuss two different 3+1 models of general relativity, namely the ADM and the BSSN, while in section \ref{numsec} we explain our numerical technique. Here we present the decomposition, the spacetime filters and the time-step control. The classical numerical testbeds are presented in section \ref{testbd}, while in section \ref{secihb} we introduce a novel technique that suppresses boundary effects, tested via gravitational wavepackets and a head-on collision of black holes. Finally, in section \ref{concl}, we summarize the main achievements of our work and discuss our conclusions.

\section{Spacetime foliation in general relativity}
\label{spacetim}
We use the same spectral algorithm for two different formulations, namely the ADM and the BSSN set of equations. In particular, we build an algorithm based on dealiased spectral methods and compare the above approaches through all the classical testbeds. As follows, we briefly summarize these two models of general relativity.

\subsection{The ADM formulation}
Gravitational fields obey to the celebrated Einstein field equation, which in vacuum can be written as
\begin{equation}
    G_{\mu \nu}=0.
    \label{einstein}
\end{equation}
Throughout this paper Latin indices will refer to spatial coordinates, running from 1 to 3, whereas Greek letters will indicate spacetime indices, spanning from 0 to 3. Although equation (\ref{einstein}) is nicely compact and elegant, is very poorly adequate for direct numerical integration \cite{ADM}. A large effort has been devoted to alternative forms of (\ref{einstein}), as for example in the case of the $3+1$ foliation of general relativity, proposed in the pioneering work by Arnowitt, Deser, and Misner \cite{Alcu}. In this case the metric is
\begin{eqnarray}
    \nonumber
    ds^2 = -\alpha^2 dt^2 +\gamma_{ij} \big(dx^i+\beta^i dt\big)\big(dx^j+\beta^j dt\big),
\end{eqnarray}
where $\alpha$, $\beta^k$ and $\gamma_{ij}$ are the lapse function, the shift vector and the spatial metric, respectively. Within the $3+1$ ADM formalism, the Einstein equations are separated into evolution equations and constraints. In any region of the spacetime, four identities must be satisfied and these are known as the Hamiltonian and the momentum constraints, namely:
\begin{eqnarray}
	\label{constraints}
        \mathcal{H} =	R+  K^2  -  K_{ij} K^{ij} = 0,\\
        \label{Hadm}
        \mathcal{M}^i= D_j\big(K^{ij} -\gamma^{ij} K\big) = 0.
\end{eqnarray}
In the above relations $K_{ij}$ is the extrinsic curvature and $K = \gamma^{ij} K_{ij}$ is its trace, while $R$ is the scalar curvature and $D_i$ is the covariant derivative. Any initial condition must obey equations (\ref{constraints})--(\ref{Hadm}) and this is known as the initial data problem \cite{baumg2, init, york1}. Associated to the above constraints, the evolution equations for the spatial metric and the extrinsic curvature are given by
\begin{eqnarray}
    	\label{eq:prima3}
	    (\partial_t - \mathcal{L}_\beta) \gamma_{ij} = -2 \alpha K_{ij},\\ 
	    \label{eq:seconda}
    	(\partial_t - \mathcal{L}_\beta) K_{ij}= \alpha \big(R_{ij}-2K_{ik}K^{k}_{~j}+KK_{ij}\big)
    	-D_i D_j \alpha, 
\end{eqnarray}
where $\mathcal{L}_\beta$ is the Lie derivative with respect to the shift vector $\beta^k$. Finally, we need the evolution equation for the lapse and the shift, choosing therefore the slicing conditions. In the framework of the Bona-Massó formalism \cite{bona1,bona3}, we choose
\begin{equation} 
    \label{slicing}
         (\partial_t - \mathcal{L}_\beta) \alpha = -\alpha^2 f(\alpha) K,
\end{equation} 
with $f(\alpha)>0$ (but arbitrary) \cite{gau4,gau}. In the case in which $f(\alpha) = 1$, the above equation describes the harmonic slicing for the lapse \cite{bona2}. In the ADM code we will set $\beta^i=0$ therefore we do not need an equation for the shift. We will introduce such an equation for $\beta^i$, in the next subsection, within the BSSN formalism.

\subsection{The BSSN formulation}
Several formulations alternative to the ADM have been proposed in the literature \cite{adj, z4}. A popular extension of the ADM is based on the work by Baumgarte and Shapiro\cite{50} and by Shibata and Nakamura \cite{baumg3}. A model known as the BSSN approach. In order to introduce the BSSN formulation, let consider first a conformal rescaling of the spatial metric of the form
\begin{equation}
    \widetilde{\gamma}_{ij} = \psi^{-4} \gamma_{ij},
    \label{tran}
\end{equation}
where $\psi$ is a conformal factor that can in principle be chosen in a number of different ways. Here we follow the approach suggested by Campanelli et al.\cite{camp}, where the conformal factor is written as $\psi^{-4} = \chi $, choosing $\chi$ such that the conformal metric $\widetilde{\gamma}_{ij}$ has unit determinant, namely $\chi = \gamma^{-1/3}$, where $\gamma$ is the determinant of the physical spatial metric. In the BSSN, the extrinsic curvature is subjected to the same conformal transformation described by equation (\ref{tran}) and its trace $K$ is evolved as an independent variable. Instead of $K_{ij}$, the new variables are $K=\gamma^{ij} K_{ij}$ and $\widetilde{A}_{ij} = \chi \Big( K_{ij} - \frac{1}{3} \gamma_{ij} K \Big)$ -- a choice that implies $tr \widetilde{A}_{ij} =0$. Finally, a new field is usually introduced, namely the  Christoffel symbols of the conformal metric $\widetilde{\Gamma}^i = \widetilde{\gamma}^{jk} \widetilde{\Gamma}^i_{jk}$. This is a crucial point since it has been proven that this choice further stabilizes numerical algorithms \cite{Alc}.

By using the above definitions and change of variables, the system of BSSN equations reads:
\begin{eqnarray}
        (\partial_t - \mathcal{L}_\beta) \widetilde{\gamma}_{ij} = - 2 \alpha \widetilde{A}_{ij},
        \label{evo1} \\ 
        \partial_t \chi = \frac{2}{3}\chi (\alpha K -  \partial_k \beta^k) +  \beta^k \partial_k \chi,
        \label{evo2} \\
        (\partial_t - \mathcal{L}_\beta) \widetilde{A}_{ij} =
        \chi \Big[ - D_i D_j \alpha + \alpha R_{ij}
        \Big]^{TF}+
        \alpha \Big(K \widetilde{A}_{ij} - 2 \widetilde{A}_{ik} \widetilde{A}^{k}_{j}\Big),
        \label{evo3}\\
        (\partial_t - \mathcal{L}_\beta)  K = - D^k D_k \alpha + \alpha \Big( \widetilde{A}_{lm}\widetilde{A}^{lm} +  \frac{1}{3} K^2 \Big),
        \label{evo4} \\ \nonumber
        \partial_t \widetilde{\Gamma}^i =  \widetilde{\gamma}^{lm} \partial_l \partial_m \beta^i + \frac{1}{3} \widetilde{\gamma}^{il} \partial_l \partial_m \beta^m  +  \beta^k \partial_k \widetilde{\Gamma}^i - \widetilde{\Gamma}^k \partial_k \beta^i + \frac{2}{3} \widetilde{\Gamma}^i \partial_k \beta^k   \\ 
         -  2 \widetilde{A}^{ik} \partial_k \alpha  
         + \alpha \Big(  2\widetilde{\Gamma}^i_{lm} \widetilde{A}^{lm}
         - \frac{3}{ \chi}\widetilde{A}^{ik} \partial_k \chi 
         - \frac{4}{3} \widetilde{\gamma}^{ik} \partial_k K  \Big).
    \label{evo5}
\end{eqnarray}
In the above set, ``$TF$'' indicates the trace-free parts of a tensor object and $D_i$ is the covariant derivative associated with the physical three-metric $\gamma_{ij}$.  The Ricci tensor associated with the physical metric is separated into two contributions, $R_{ij} = \widetilde{R}_{ij} + R^{\chi}_{ij}$, where 
\begin{eqnarray}
\fl ~~~~~~~~~~~~\widetilde{R}_{ij}= - \frac{1}{2} \widetilde{\gamma}^{lm} \partial_l \partial_m   
    \widetilde{\gamma}_{ij}
    + \widetilde{\gamma}_{k(i} \partial_{j)} \widetilde{\Gamma}^k 
    + \widetilde{\Gamma}^k  \widetilde{\Gamma}_{(ij)k}  
     + \widetilde{\gamma}^{lm}  \Big( 2 \widetilde{\Gamma}^k_{l(i}  \widetilde{\Gamma}_{j)km} 
    + \widetilde{\Gamma}^k_{im}  \widetilde{\Gamma}_{klj} \Big)
    \label{evo6}
\end{eqnarray}
is the Ricci tensor related to the conformal metric and 
\begin{eqnarray} 
  \fl  ~R_{ij}^\chi =  \frac{1}{2 \chi} \Bigg\{ 
    \bigg[
    \partial_i \partial_j \chi
   - \frac{\partial_i \chi \, \partial_j \chi}{2 \chi}
   - \widetilde{\Gamma}^k_{ij} \partial_k \chi  \bigg]
   + \widetilde{\gamma}_{ij} \bigg[
    \widetilde{\gamma}^{lm} \bigg( 
    \partial_l \partial_m \chi 
    - \frac{3 \, \partial_l \chi \, \partial_m \chi}{2 \chi}   \bigg)
    - \widetilde{\Gamma}^k \partial_k \chi
    \bigg]
    \Bigg\}
\label{evo7}
\end{eqnarray}
denotes the additional term that depends on $\chi$. In addition to equation (\ref{slicing}), we use the following slicing conditions \cite{book}:
\begin{eqnarray}
(\partial_t - \mathcal{L}_\beta) \beta^i &=& \frac{3}{4} B^i ,\label{slic2} \\
(\partial_t - \mathcal{L}_\beta) B^i &=& \partial_t \widetilde{\Gamma}^i - \beta^j \partial_j  \widetilde{\Gamma}^i - \eta B^i.
\label{slic3}
\end{eqnarray}
Here the factor $3/4$ is somewhat arbitrary, but leads to good numerical results and the parameter $\eta$ is a positive constant \cite{baumg1}. In all the evolution equations above, $\widetilde{\Gamma}^i$ is replaced by $-\partial_j \widetilde{\gamma}^{ij}$ wherever it is not differentiated.

In addition to the evolution equations, the BSSN variables must also obey to the constraints in equations (\ref{constraints})-(\ref{Hadm}), that now become
\begin{eqnarray}
    \nonumber
    \mathcal{H}= R - \widetilde{A}_{lm} \widetilde{A}^{lm} + \frac{2}{3} K^2=0,\\
    \nonumber
    \mathcal{M}^i=\partial_k \widetilde{A}^{ik} + \widetilde{\Gamma}^i_{lm} \widetilde{A}^{lm}- \frac{3}{2 \chi } \widetilde{A}^{ik} \partial_k \chi - \frac{2}{3} \widetilde{\gamma}^{ik} \partial_k K =0.
\end{eqnarray}
The specific choice of evolution variables introduces five additional constraints, namely:
\begin{eqnarray} 
det \, \widetilde{\gamma}_{ij} = 1,
\label{cons1} \\
tr \widetilde{A}_{ij} = 0,
\label{cons2} \\
\mathcal{G}^i=  \widetilde{\Gamma}^i + \partial_j \widetilde{\gamma}^{ij}=0.
\end{eqnarray}
During the numerical simulations, we enforce the algebraic constraints in equations (\ref{cons1}) and (\ref{cons2}), as discussed in \cite{campan}. The remaining constraints, $\mathcal{H}$, $\mathcal{M}^i$ and $\mathcal{G}^i$ are not actively enforced and are used to monitor the accuracy of the numerical solutions. For more details on these additional  constraints, see for example Ref.\cite{Alc}.  As a summary, we solve equations (\ref{evo1})-(\ref{evo5}) for the new variables, by decomposing the Ricci tensor with (\ref{evo6})-(\ref{evo7}) and closing the system for the lapse $\alpha$ and the shift $\beta^i$ via the slicing equations (\ref{slicing}), (\ref{slic2}) and (\ref{slic3}).

\reply{
    Finally, it is worth mentioning also the generalized harmonic formulation, for example used by spectral or other efficient finite difference codes \cite{HAD}. This latter formulation is shown to be symmetric hyperbolic and admits well posed numerical schemes, being in fact the first method to achieve stable evolutions of more than a single orbit, as discussed in the work by Pretorious \cite{PretoriousB2005}.
}

\section{The numerical technique}
\label{numsec}
In order to solve the gravitational fields equations, using both the ADM and the BSSN formalism, we make use of spectral techniques \cite{canuto, FFFT, orszag2, orszag3, spectral}. More specifically, in our code we use the {\it pseudo-spectral} method, which is a technique characterized by a recurrent switch, back-and-forth, between the physical (real) and the Fourier (complex) space. The rule is quite simple: derivatives are computed in the spectral space while products are computed in the physical space (in order to avoid convolutions.) This iterative transformation, as we will see in the next subsection, needs to be regularized via a dealiasing procedure. Classical spectral and pseudo-spectral methods are widely used for complex problems that involve strong couplings and nonlinearities. The main advantage of these techniques relies on the fact that each field and all its derivatives are  represented at the collocation points in a very accurate way. This leads to very precise solutions, with an error on the computation of derivatives that is on the order of the machine truncation error. 

In order to compute any spatial derivative, imposing periodic boundary conditions in all the three spatial dimensions, our code makes use of standard Fast Fourier Transforms (FFTs). More specifically, each field $f({\bf x}, t)$ is projected on a 3D spatial box, over a lattice with a number of meshes ${\bf N} = (N_x, N_y, N_z)$. We decompose the ADM and the BSSN fields as:
\begin{eqnarray}
    f_{\bf N}(\bf x,\it t) = \sum _{{\bf k}} \widetilde{f}_{\bf k}(t) ~e^{i  {\bf k} \cdot {\bf x}}, 
\label{fou}
\end{eqnarray}
with $f_{\bf N}$ being the fields of the 3+1 foliation (i.e. $\gamma_{ij}, K, K_{ij}$ and so
on.), at the grid points. 
\reply{ 
As usual, $\widetilde{f}_{\bf k}(t) \in \mathbb{C}$ are the Fourier coefficients
}
and ${\bf k}= (k_x, k_y, k_z)$ is the $k$-vector. In each Cartesian direction, $k=\frac{2\pi}{L_{0}}m$, $L_0$ is the box length, $m=0, \pm 1, \pm 2, \dots \pm N_k$ with $N_k=N/2$ being the Nyquist mode and $N$ is the number of points along this direction.

Even if pseudo-spectral methods are among the most precise numerical techniques, particularly suitable for solving small scale complex problems such as wave dynamics, turbulence and so on, great care must be dedicated to the computation of nonlinear terms. Nonlinearity characterizes gravitation, as in the case of hydrodynamics and plasma-dynamics (although in these cases the level of nonlinearity is certainly lower, as we shall discuss later.) In the spectral methods representation, the nonlinear terms become a convolution and there are several transform-based techniques for evaluating it efficiently \cite{canuto, ggg, FFT}.  Numerical problems might arise from the so-called aliasing instabilities \cite{gal}, since any product among fields creates new $k$-components (spectral leakage). In discrete Fourier transforms, alias happens when the size $N_k$ of sampled Fourier modes is less than the effective size of modes. In such cases, due to the cyclic assumption in discrete Fourier transform, a Fourier mode with wavenumber out of the size-range is aliased to another wavenumber in the domain. In a time-evolving system, this aliasing creates numerical instabilities. The importance of eliminating the aliasing error (dealiasing) has been studied since pioneering works by Orszag et al. \cite{orszag}.  As follows, we briefly mention this problem, well-known in hydrodynamics, and propose a similar ``cure'' to this numerical instability in the case of gravitational dynamics.

Let consider the simplest possible nonlinearity, i.e. the product of two functions $f(x)$ and $g(x)$ in a 1D, periodic domain (the 3D extension is straightforward). The truncated Fourier series with a set of $\{N_k\}$ modes are
\begin{eqnarray}
\nonumber
    f_N(x) =  \sum _{p \in \{N_k\}} \widetilde{f}_p~ e^{i p x},~~~~~~g_N(x) =  \sum _{q \in \{N_k\}} ^{m} \widetilde{g}_q~ e^{iq x},
\end{eqnarray}
where $\widetilde{f}_p$ and $\widetilde{g}_q$ are the complex Fourier coefficients, as defined in (\ref{fou}). The product of the above two functions $Q(x)=f(x)g(x)$ becomes a convolution product in Fourier space, namely
\begin{eqnarray}
    \widetilde{Q}_k = \int \sum_p \sum_q \widetilde{f}_p~\widetilde{g}_q ~ e^{i (p+q-k) x} d x = \sum_{k=p+q} \widetilde{f}_p~\widetilde{g}_q,
\label{prod2}
\end{eqnarray}
where the last sum is the convolution on all the possible modes $p$ and $q$ of the set $\{N_k\}$ that satisfies the resonance $k=p+q$. It can be clearly seen that the product (\ref{prod2}) generates high order harmonics with respect to the Fourier series truncated at $N_k$. They will contribute to the well-known \textit{aliasing error} \cite{Boyd}. The dealiasing technique consists of making the maximum number of Fourier modes $k^*$ in the above centered convolution free of aliasing, by extending the vector on both sides by zeros, to a larger vector of size $N_k$ (zero padding). Analogously, it is easy to demonstrate that the full dealiasing can be achieved by defining a $k^*$ in such a way that  all coefficients with $p,q,...>k^*$ are zero. For a quadratic nonlinearity like in equation (\ref{prod2}), it has been demonstrated that is sufficient to filter out modes with $k> k^*= 2 N_k/3 = N/3$. This fully eliminates the aliasing instability \cite{canuto} and is often called the $2/3~ rule$, since the number of dealiased modes is 2/3 of the total number of modes.

Compared to the most widely inspected cases, such as the Burgers equations and Navier-Stokes, gravitation involves high-order products and derivatives. In particular, from a quick analysis of the ADM equations, it is clear that one has products of order $p=4$, or more. The nonlinearity is not completely well-defined because of contravariant tensors which  complicate the degree of the nonlinearity, involving inverse computation of the metric. In these cases it is still possible to suppress the aliasing instability by finding an accurate truncation \cite{gal}, which establishes a direct comparison between spectral methods and a full Galerking truncation. The above quadratic dealiasing described before can be extended to simple polynomial nonlinear terms of order $p$, by setting $k^*= \frac{N}{1+p}$, therefore reducing the number of degrees of freedom in the Fourier space.

In this work we will inspect the role of the aliasing truncation and also the shape of the filter \cite{gal2, TaylorG}. The technique is very simple to implement: on the final product it is enough to set ${\widetilde Q}_k = 0$ for $k>k_\star$. In the case of the Einstein field equations, it is important to consider the very high nonlinearity of the system, where one has products of the type:
\begin{eqnarray}
\nonumber
    \partial_t K_{xx} \sim ....+ \frac{1}{4}\alpha \gamma^{yx}(\partial_y \gamma_{xx})\gamma^{yy}(\partial_x\gamma_{yy}) + ...
\end{eqnarray}
The above quantity has a nonlinearity of order 5, which corresponds to a challenging convolution in the Fourier space. It is easy to envision a process in which the above products produce immediately high-order harmonics and therefore a pronounced aliasing instability. 
\begin{figure}
\centering
\includegraphics[width=100mm]{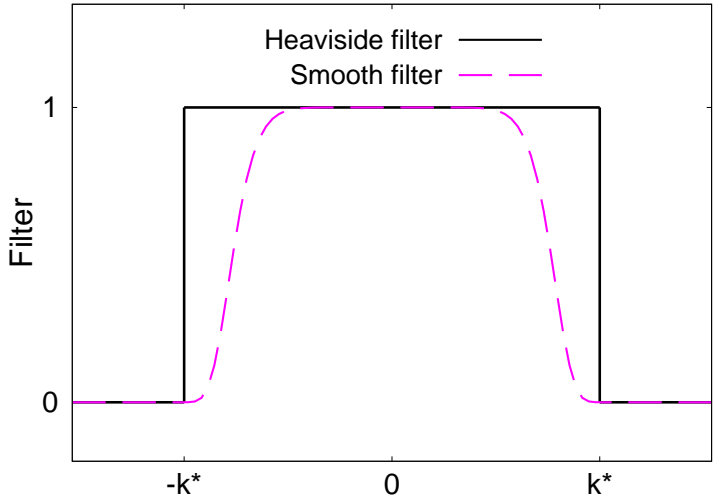}
\caption[ ]{\footnotesize{} Comparison of the two anti-aliasing filters described by equations (\ref{filteralias2}) and (\ref{filteralias1}). The figure shows a section of the Fourier space, where all the $ |k| \geq k^*$ are truncated. The Heaviside filter is a unit step-function, while the smooth filter goes from 0 to 1 more gently. In the example it has been used $a = 10$ to emphasize the smoothness of this filter. In all our simulations we set $a=20$.
}
\label{filtri}
\end{figure}
In order to suppress the above numerical pathology, different values of $k^*$ have been chosen, depending on the difficulty of the simulation and on the initial data. Generally, by filtering high Fourier modes, the price to pay is the loss of effective resolution (information). However, by suppressing these high-$k$'s couplings, the codes become more stable and accurate.

Apart of the mobile cutoff at $k^*$, in this paper we propose two different shapes of the spectral filter, described as follows. For each representation, we define
\begin{equation}
    f_{\bf N}(\bf x,\it t) = \sum _{{\bf k}} \widetilde{f}_{\bf k}(t) ~e^{i  {\bf k} \cdot {\bf x}}  \Phi_{k^*}(\bf k),
\label{fltr}
\end{equation}
where $\Phi_{k^*}({\bf k})$ is the spectral anti-aliasing filter. The first type, is the classical spherical truncation technique, namely a ``Heaviside'' filter (H) defined as:
\begin{equation}
	\Phi_{k^*}^{H}(\bf k)= \cases{1&for $|{\bf k}|\leq k^*$,\\
	0&for $|{\bf k}|> k^*$.}
	\label{filteralias2}
\end{equation}
A second smooth filter (S), usually adopted for treating discontinuities and singularity formation \cite{TaylorG, Lele}, is given by
\begin{equation}
	\Phi_{k^*}^{S}({\vec k})= \exp(-a \xi^a),
	\label{filteralias1}
\end{equation}
where $\xi = |\vec k|/k^*$ and $a$ is a free parameter. It's clear that in the limit $a \rightarrow \infty$, the two filters become identical. A cartoon of the two filters is represented in figure \ref{filtri}. Generally, in our numerical experiments, we will use $k^* = \infty$ (no-filter), $k^*=N/2$ (grid-size), $k^*=N/3$ (typical quadratic nonlinearities),  $N/4$ and so on. For the smooth filters $\Phi_{k^*}^{S}(k)$ in equation (\ref{filteralias1}), we set $a=20$.

\reply{
    The main challenge with using spectral methods for numerical relativity is the fact that they require sufficiently smooth solutions. A spectral decomposition can provide derivative evaluations with exponential accuracy (that is, higher than any polynomial power of the grid spacing) only if the data being represented is $C_\infty$. When derivatives are continuous at some lower order, then the error might rise too much. The filtering techniques might help to suppress the above problem. However, the accuracy of the method needs to be verified via a convergence analysis, especially when evolving non-smooth solutions. These numerical tests are reported in the Appendix.
}

Note that even flux-conservative, Godunov-type schemes, although they formally solve the ideal equations, retain a considerable amount of artificial dissipation in the flux-reconstruction scheme \cite{filter1, filter2}. The dissipation of these space-filters is sometimes difficult to quantify. In our much simple pseudo-spectral approach, the algorithm is stabilized by definite spectral kernels, that act only at very small scales, over a controlled and limited band-width of modes.

The above pseudo-spectral technique, used for the products and the spatial derivatives, has been associated to a simple 
\reply{
    second-order
} 
Runge--Kutta method for the time integration \cite{nrecipes}. At each time discretization $n$, the function is advanced in time by using 
\begin{equation}
    f^{n+1}=f^{n}+\Delta t \,F\left[t^n+\frac{\Delta t}{2}, f^{n} + \frac{\Delta t}{2} F(t^n, f^{n})\right], 
\label{rk2}
\end{equation}
where $f^{n}\equiv f(x, y, z, t^n)$ is any of the gravitational fields at the time $t^n=n \Delta t$, $F$ is the time-derivative defined by the 3+1 governing equations, and  $\Delta t$ is the time-step, chosen to be sufficiently small. 
\reply{
    It is worth saying that second-order evolution might be too dissipative, especially in very stressful conditions such as the inspiraling of binary systems, while it might still manifest a good ``convergence''. In the Appendix, we report on the adaptation of the current method to fourth-order Runge-Kutta, which is surely more precise, but is more numerically expensive. Moreover, in a pseudo-spectral algorithm, where the number of allocated arrays is large, memory-saving becomes a challenge. Hereafter, we will therefore show the main results for the second-order scheme, which has the advantage of being simpler, more dissipative for the highest frequencies and adaptable to other algorithms such as the implicit Crank-Nicholson (presented in Section \ref{secihb}). In our integration scheme, the time-step can be changed during the evolution, with a simple predictor scheme, as described in the following subsection.
}

\subsection{A control on the stability of simulations}
\label{rsc1}
In numerical methods, the Courant-Friedrichs-Lewy (CFL) condition 
\reply{
is a necessary stability condition 
}
for the solution of certain partial differential equations problems \cite{CFL}. In the simplest case of the motion of a wave traveling through a discrete spatial grid with speed $v$, the CFL condition imposes $\Delta t < C \Delta x/v$, namely a condition on the characteristic time of integration. In the above condition, $C$ is the Courant number and is in general a number smaller than one. Numerical simulations of purely hyperbolic systems must satisfy this condition.  The case of numerical relativity is more complicated because of the strong nonlinear couplings  and because of the low level of hyperbolicity of the $3+1$ models (the strictly--hyperbolic nature of $3+1$ models is still an open issue \cite{shock_capturing}). The above complications might lead to stability conditions that are more restrictive than a simple CFL. A general criterion for the determination of the time-step integration $\Delta t$ is therefore necessary in order to have a good balance between time-efficiency of the integration and avoidance of nonphysical solutions. Here we re-elaborate the CFL idea, in a more general sense, as follows. 

For any ADM or BSSN dynamic field $f$, one can estimate a global characteristic time as
\begin{equation}
    \mathcal{T}_f = \frac{ \delta f }{\sqrt{ \Big \langle \left( \frac{\partial f}{\partial t}\right)^2} \Big \rangle}, 
\label{dyntime}
\end{equation}
where $\delta f=\sqrt{\langle \left(f -\langle f \rangle\right)^2 \rangle}$ is the $rms$ of the field and $\langle \dots \rangle$ indicates a volume average. The above quantity can be interpreted as a typical dynamical time of $f$, evaluated as the ratio between the level of fluctuations and its typical time-derivative, similarly to plasma characteristic times in turbulence \cite{nature}.  The characteristic time in equation (\ref{dyntime})  is defined at any given time of the foliation and does not involve any complicated computation in explicit time-integration schemes. For any dynamical field $f$, one has 
$
\frac{\partial f}{\partial t}\simeq \frac{\Delta f}{\Delta t}
$
which, from the time discretization in equation (\ref{rk2}), corresponds to $F\left[t^n+\frac{\Delta t}{2}, f^{n} + \frac{\Delta t}{2} F(t^n, f^{n})\right]$. From the latter approximation, at the intermediate step, for example, one can estimate the dynamical time in the equation (\ref{dyntime}). Note that one can also build local versions of this time, or can measure the minimum time by finding the maximum of the denominator. 

In analogy with the CFL condition, it is possible to use the characteristic time $\mathcal{T}_f$ for guessing an appropriate $\Delta t$. If the system undergoes fast evolution, $\mathcal{T}_f$ will drop off and the algorithm can self-adjust $\Delta t$. This might indicate a ``precursor'' of a simulation blow-up, as we shall see in our numerical experiments. Throughout this work, we will introduce the {\it Running Stability-Check} (RSC), which is a method that consists in a continuous adjustment of the Runge-Kutta time-step, based on the measure of $\mathcal{T}_f$ in equation (\ref{dyntime}), such that 
\begin{equation}
    \Delta t < C~\min_f \{ \mathcal{T}_f\}.
\label{rscc}
\end{equation}
Here $C$ is similar to the Courant number. The method simply corresponds to a \textit{adaptive time-refinement}, typical of many schemes such as the leapfrog integration, except here is constrained with the minimum characteristic time given by (\ref{dyntime}). 
\reply{
    This strategy is very similar to the adaptive stepsize control for Runge-Kutta, described in standard numerical textbooks \cite{nrecipes}. The general purpose of these classical methods is the achievement of some predetermined accuracy in the solution, with minimum computational effort. Here we re-elaborate this technique, but instead of establishing a desired error, we define a global time $\mathcal{T}_f$ for the system and keep the integration step always smaller than this dynamical time.
} 
Note also that the above technique might provide a general condition not only in the case of Runge-Kutta methods, but in any explicit time-integration scheme.

\section{Standard numerical testbeds}
\label{testbd}
In this section we will perform direct numerical simulations in order to validate our numerical procedure, via standard gravitational testbeds \cite{Alc, Dum, bab2, baumg2}. For each initial data, we will check the accuracy by inspecting the conservation of several quantities, for both the ADM and BSSN formalism. Regarding the spatial integration, we will show how to stabilize the pseudo-spectral code by varying our dealiasing filter in the projection (\ref{fltr}), by changing both the $k^*$  and the filter shape. For the time integration, in each test, will apply the RSC by monitoring the characteristic times in (\ref{dyntime}). After these typical, fundamental tests, we will show a new possible initial data that leads to stationary, large amplitude, nonlinear waves. These waves bounce with growing amplitude and therefore represent another good test for numerical relativity. We will then move to more stressful conditions, such as the Gowdy spacetime (both forward and backward in time) and the Schwarzschild equilibria. All the simulations performed are summarized in table~\ref{table}, where we report all the main parameters used, such the type of the initial condition, the amplitude of the  perturbations, the number of points of the spatial grid, the $k^*$ of anti-aliasing filter, the type of filter, the formalism adopted, and so on. In the last column, for the reader's convenience, we also report the stability of the run.

\begin{table}
\caption{\label{table} Table of simulations. From left to right: the run number, the initial condition, the amplitude of the perturbation, the mesh points, the $k^*$ of the filter, the filter type, the RSC status, the IBH status, the formalism used and, finally, the stability of the simulation.} 
{\def\arraystretch{1.02}\tabcolsep=2.5pt
\begin{tabular}{@{}llllllllll}
\br
 RUN & IC type             & $A$                   & $ N $   &  $k^*$   & Filter & RSC & IHB & Formalism & Stable \\
\mr
1   & Gauge wave           & $0.01$                & $64$    & $N/3$    & S & off & off & ADM  & no \\ 
2   & Gauge wave           & $0.01$                & $64$    & $N/3$    & S & on  & off & ADM  & yes \\ 
3   & Gauge wave           & $0.9$                 & $128$   & $\infty$ & / & on  & off & ADM  & no \\
4   & Gauge wave           & $0.9$                 & $128$   & $N/3$    & S & on  & off & ADM  & yes \\ 
5   & Gauge wave           & $0.96$                & $128$   & $N/2.5$  & H & on  & off & ADM  & yes \\   
6   & Gauge wave           & $0.96$                & $128$   & $N/2.5$  & S & on  & off & ADM  & yes \\ 
7   & Gauge wave           & $0.01$                & $128$   & $N/3$    & S & on  & off & BSSN & yes \\ 
8   & Gauge wave           & $0.1$                 & $128$   & $N/3$    & S & on  & off & BSSN & yes \\ 
9   & Robust stab. (n = 1) & $|10^{-10}|/\rho_1^2$ & $64$    & $\infty$ & / & off & off & ADM  & no \\  
10  & Robust stab. (n = 2) & $|10^{-10}|/\rho_2^2$ & $128$   & $\infty$ & / & off & off & ADM  & no \\ 
11  & Robust stab. (n = 4) & $|10^{-10}|/\rho_4^2$ & $256$   & $\infty$ & / & off & off & ADM  & no \\ 
12  & Robust stab. (n = 1) & $|10^{-10}|/\rho_1^2$ & $64$    & $N/4 $   & H & off & off & ADM  & no \\ 
13  & Robust stab. (n = 1) & $|10^{-10}|/\rho_1^2$ & $64$    & $N/4 $   & S & off & off & ADM  & yes \\ 
14  & Robust stab. (n = 1) & $|10^{-10}|/\rho_1^2$ & $64$    & $N/4 $   & H & off & off & BSSN & no \\ 
15  & Robust stab. (n = 1) & $|10^{-10}|/\rho_1^2$ & $64$    & $N/4 $   & S & off & off & BSSN & yes \\ 
16  & Robust stab. (n = 1) & $|10^{-10}|/\rho_1^2$ & $64$    & $N/4 $   & H & on  & off & BSSN & yes \\ 
17  & Robust stab. (n = 1) & $|10^{-10}|/\rho_1^2$ & $64$    & $N/4 $   & S & on  & off & BSSN & yes \\ 
18  & Standing wave        & $0.1$                 & $64$    & $\infty$ & / & on  & off & ADM  & no \\ 
19  & Standing wave        & $0.1$                 & $64$    & $N/3$    & S & on  & off & ADM  & yes \\ 
20  & Standing wave        & $8$                   & $64$    & $N/3$    & H & on  & off & ADM  & no \\
21  & Standing wave        & $8$                   & $64$    & $N/3$    & S & on  & off & ADM  & yes \\
22  & Gravitational wave   & $10^{-8}$             & $64$    & $N/4$    & S & off & off & ADM  & yes \\
23  & Gravitational wave   & $10^{-8}$             & $128$   & $N/8$    & S & off & off & BSSN & yes \\
24  & Gowdy forw. (n = 1)  & /                     & $64$    & $N/5$    & S & on  & off & BSSN & yes \\ 
25  & Gowdy forw. (n = 2)  & /                     & $128$    & $N/5$    & S & on  & off & BSSN & yes \\ 
26  & Gowdy forw. (n = 4)  & /                     & $256$    & $N/5$    & S & on  & off & BSSN & yes \\ 
27  & Gowdy backward       & /                     & $64$    & $N/5$    & S & on  & off & ADM  & yes \\ 
28  & Gowdy backward       & /                     & $64$    & $N/5$    & S & on  & off & BSSN & yes \\ 
29  & Schwarzschild BH     & /                     & $64^3$  & $N/2$    & H & on  & off & ADM  & no \\ 
30  & Schwarzschild BH     & /                     & $64^3$  & $N/2$    & S & on  & off & ADM  & no \\
31  & Schwarzschild BH     & /                     & $64^3$  & $N/2$    & H & on  & off & BSSN & no \\ 
32  & Schwarzschild BH     & /                     & $64^3$  & $N/2$    & S & on  & off & BSSN & yes \\ 
33  & GW absorbtion        & $10^{-5}$             & $128$   & $N/4$    & S & on  & on  & BSSN  & yes \\ 
34  & Head on collision    & /                     & $128^3$ & $N/2$    & S & on  & off & BSSN  & no \\ 
35  & Head on collision    & /                     & $128^3$ & $N/2$    & S & on  & on  & BSSN & yes \\ 
\br
\end{tabular} }
\end{table}

\subsection{The gauge wave}
\label{ggwave}
The gauge wave test is based on a gauge transformation of the Minkowski spacetime. As suggested by Ref.~\cite{Alc}, the metric is given by
\begin{equation}
    ds^2 = -H(x,t)\, dt^2 +H(x,t) \,dx^2 +dy^2+dz^2, 
\label{wavetest1}
\end{equation}
where $H(x,t) = 1-A~\sin\big[2 \pi (x-t)\big]$ describes a sinusoidal modulation of amplitude $A$, propagating along the $x$-axis. Since derivatives are zero in the $y$ and $z$ directions, the problem is essentially 1D. The metric in equation (\ref{wavetest1}) implies $\beta^i=0$, and $K_{xx}=-\partial_t \gamma_{xx} / 2 \alpha$. For the spatial metric $\gamma_{ij}$ one gets
\begin{eqnarray}
\nonumber
    \gamma_{ij}=
    \left( \matrix{ 1-A~\sin[2 \pi (x-t)] & 0 & 0 \cr
0 & 1 & 0 \cr
0 & 0 & 1 \cr} \right),
\end{eqnarray}
while the only nontrivial component of the extrinsic curvature is
\begin{eqnarray}
\nonumber
    K_{xx}=-A\pi\frac{ \cos[2 \pi (x-t)]}{\sqrt{1-A~\sin[2 \pi (x-t)]}}.
\end{eqnarray}
It is easy to demonstrate that the above fields satisfy the initial data constraints in equations (\ref{constraints}) and (\ref{Hadm}).

In order to stress the stability of our new code, we solved the ADM equations (\ref{eq:prima3})--(\ref{slicing}), with both small- and large-amplitude initial conditions \cite{Dum}. We chose a spatial domain $x \in [ 0, 1 ]$ with a number of meshes $N_x = 64$. The initial time-step is $\Delta t = 5 \times 10^{-3}$ and, according to the literature, we used the harmonic slicing, namely $f(\alpha)=1$ in equation (\ref{eq:prima3}). For the first run, RUN$_1$ in table~\ref{table}, we impose $A=10^{-2}$ and for the numerical technique we choose the filtering mode $k^*=N_x/3$, with the smooth filter described by equation (\ref{filteralias1}). In order to test the RSC method described in section \ref{rsc1}, the run was performed twice: with a fixed time-step (RUN$_1$) and with an adaptive-time-refinement suggested by the RSC method (RUN$_2$).
\begin{figure}
\hspace{-8pt}
\includegraphics[height=62mm,width=160mm]{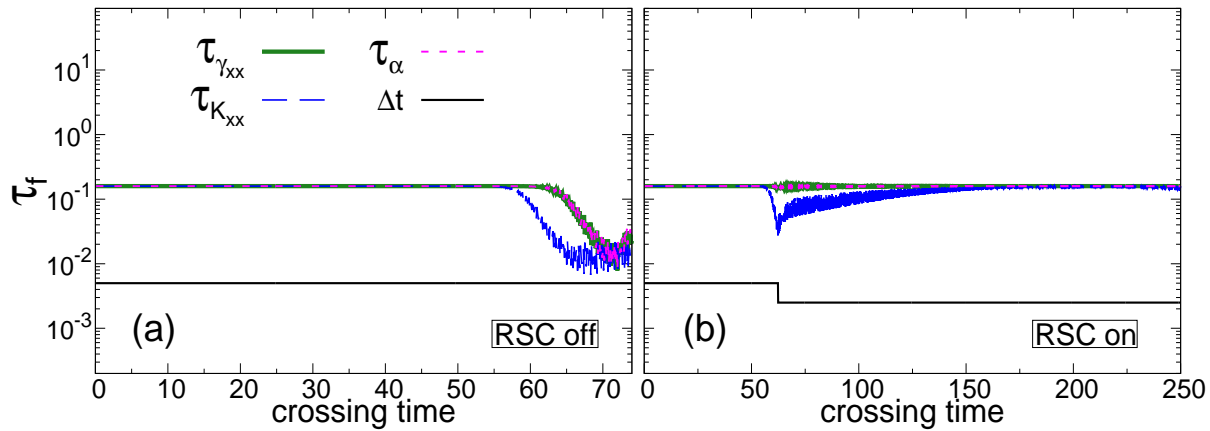}
\caption[RSC conditions and adaptive time refinement]
{\footnotesize{}Small amplitude gauge wave test for the ADM code. Left: RSC conditions without adaptive time refinement (RUN$_1$). The code crashes at $t\simeq 75$. Right: RSC conditions with $C = 1/6$ (RUN$_{2}$): the simulation remains stable. The full black line represents the time-step.}
\label{a1}
\end{figure}
Figure \ref{a1} (a) shows the evolution of the characteristic times given by equation (\ref{dyntime}). Note that only the characteristic times associated to the non-null fields have been reported, namely $\mathcal{T}_{K_{xx}}$, $\mathcal{T}_{\gamma_{xx}}$ and $\mathcal{T}_{\alpha}$. The full (black) line represents the time-step $\Delta t$ of the simulation. One can immediately see that when a dynamical time $\mathcal{T}_f$ becomes on the order of the time-step, instabilities arise and the code crashes. In the second test, reported in the panel (b) of the same figure, we repeated the experiment with the RSC switched on. In particular, we set the adaptive time refinement in equation (\ref{rscc}) with $C=1/6$, and the time-step has been halved any time that the condition is violated. The run is therefore stabilized and carried out until 250 crossing times. This simulation is reported as RUN$_{2}$ in table~\ref{table}. Similar behavior has been obtained for other initial data (not shown here).

\begin{figure} 
\hspace{-15pt}
\includegraphics[height=110mm,width=162mm]{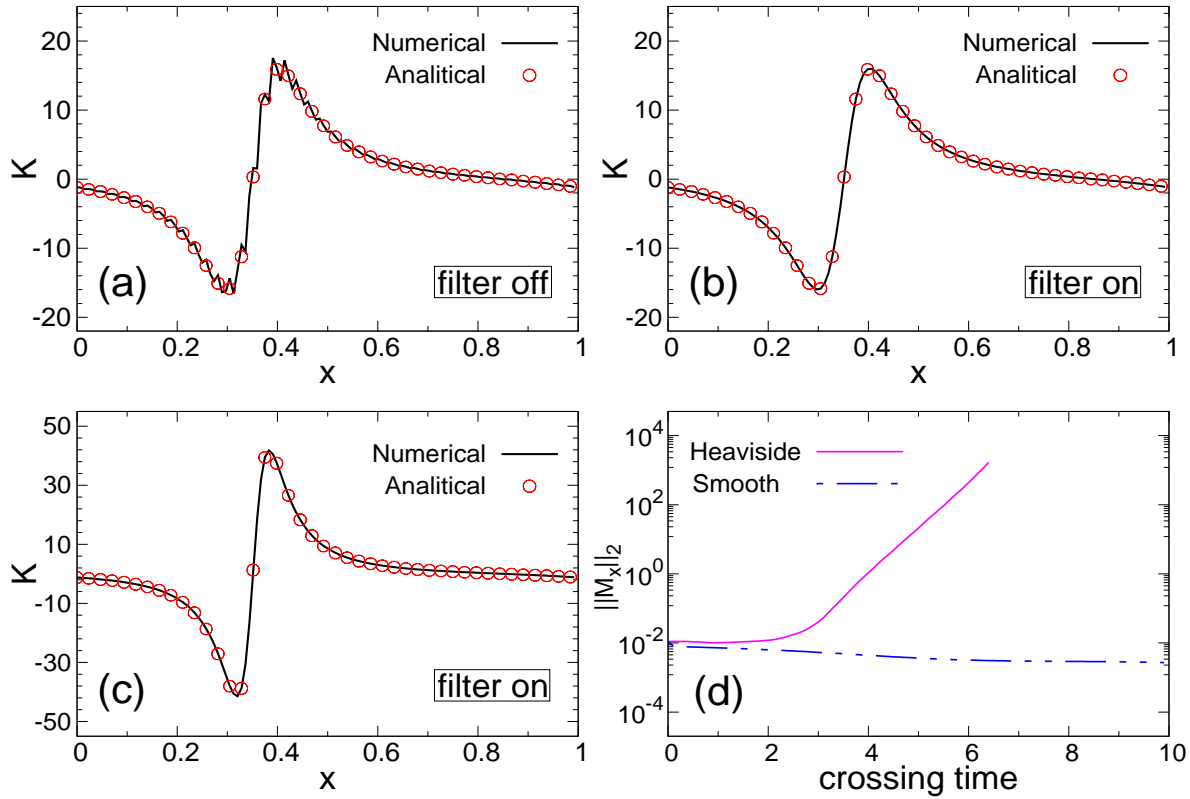}
\caption[]{\footnotesize{} 
Gauge wave test with the ADM code. (a) RUN$_{3}$, the trace of the extrinsic curvature $K$ in nonlinear regime ($A=0.9$), without filter, at $t=1.2$ (full black line). The exact solution is reported with (red) circles. (b) RUN$_{4}$, same as RUN$_3$ but with a smooth filter at $k^* = N_x/3$. The anti-aliasing filter leads to an improved wave-form. (c) RUN$_{6}$, $K$ for the case with very high amplitude $A=0.96$, at $t=10$, using the smooth filter. (d) Evolution of $x-$momentum constraint for the very high amplitude case, using Heaviside (RUN$_5$) and smooth (RUN$_6$) filter.} 
\label{ap}
\end{figure}

For the large-amplitude regime of the gauge-wave test, we have chosen $A=0.9$ and $A=0.96$, with a spatial grid along the $x$-axis of $N_x=128$ points and $\Delta t=10^{-3}$. The other parameters are the same as in the small amplitude case, including the slicing gauge. Again we have performed two experiments: first without anti-aliasing filter (RUN$_3$) and second with a $k^* = N/3$ smooth filter (RUN$_4$), defined in equation (\ref{filteralias1}). A comparison between the waveforms of the extrinsic curvature trace $K$, in both cases, has been represented in the top-row of figure \ref{ap}. As it can be seen, there is a net improvement with the anti-aliasing filter. One can observe, overall, an excellent agreement between the exact and the numerical solutions. In the unfiltered case numerical instabilities arise, as can be seen in panel (a). The analogous filtered simulation is reported in (b).

A second, more challenging test has been performed, with a larger perturbation, i. e. $A=0.96$. We found that now a stable solution can be achieved even with $k^* = N/2.5$. In order to test the different anti-aliasing filters we performed the test twice: RUN$_{5}$ uses the Heaviside filter defined in equation (\ref{filteralias2}) while RUN$_{6}$ adopts the smooth filter (\ref{filteralias1}). The panel (d) of figure \ref{ap} shows the $L_2$ errors of the $x-$momentum constraint, where $L_2$ norm is given by
\begin{equation}
    \Vert{\dots}\Vert_2=\sqrt{\frac{\int_{\Omega} \left(\dots \right)^2\, d\Omega\,\sqrt{|\gamma|}}{\int_{\Omega}d\Omega\,\sqrt{|\gamma|}}},
\label{norml2}
\end{equation}
where $d\Omega\,\sqrt{|\gamma|}$ is the volume element. The smooth filter again stabilizes the code. In figure \ref{ap} (c) we report the waveform of $K$ at time $t=10$, which is in very good agreement with the exact solution. It is important to emphasize that the last set of runs, where $A=0.96$, is considered to be a very strong stress for codes, as discussed in \cite{Dum}.
\begin{figure}[ht]
\hspace{-5pt}
\includegraphics[height=62mm,width=160mm]{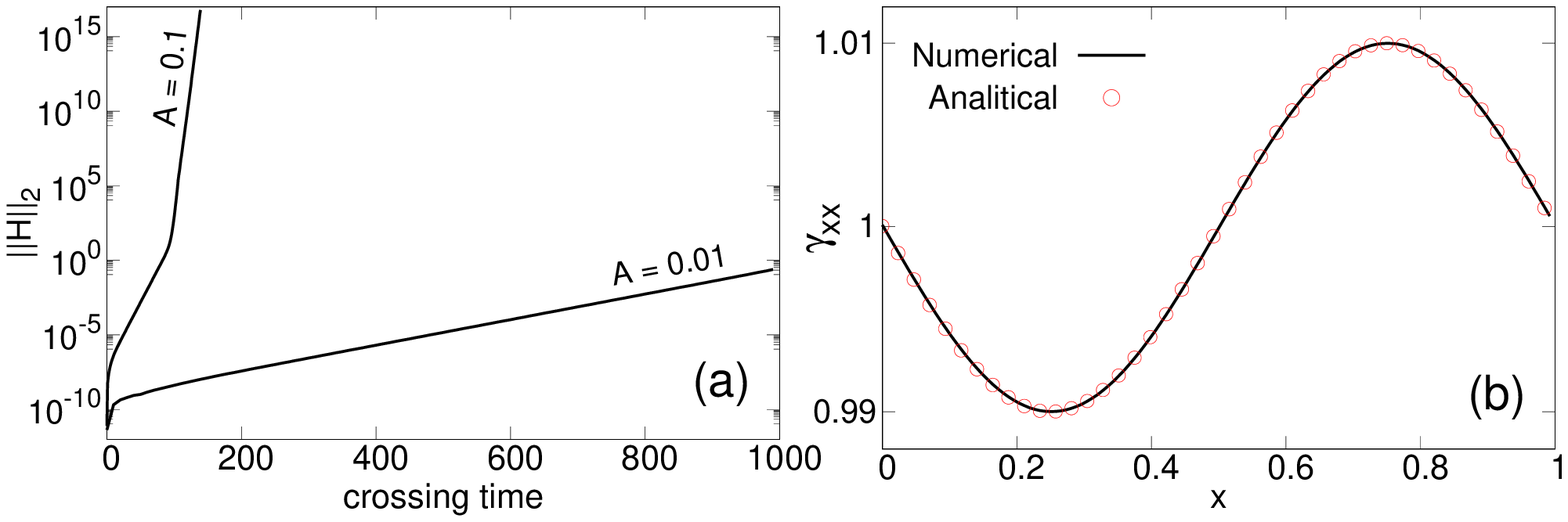}
\caption[ ]{\footnotesize{} (a) Hamiltonian violation $||H||_2$ for the gauge wave, for the BSSN simulations RUN$_7$ ($A=0.01$) and RUN$_8$ ($A=0.1$). These results are in good agreement with Ref.~\cite{campan, daverio}. (b) $\gamma_{xx}$ at $t=500$, for RUN$_7$ (line) and analytical (circles).}
\label{gaugebssn}
\end{figure}

For this testebed, we obtained similar results for the BSSN code. Our conformal model follows the approach by Zlochower et al. \cite{campan}. We used $A=10^{-2}$ (RUN$_7$) and $A=10^{-1}$ (RUN$_8$), a spatial grid with $N_x = 128$ mesh points and a time-step $\Delta t=5 \times 10^{-4}$. The harmonic gauge in equation (\ref{slicing}) has been used. In order to better stabilize the simulation, a smooth filter with $k^* = N/3$ has been chosen, for both amplitudes, similarly to the ADM cases.  The left panel of figure \ref{gaugebssn} shows the $L_2$ errors of the Hamiltonian for the two amplitudes, while in the right panel we compare the wave-form of $\gamma_{xx}$, at $t=500$. At this time, the norm of the Hamiltonian violation $||H||_2$ is about $0.1 \%$ of A, for the case with $A=10^{-2}$. The run shows an excellent agreement between numerical and exact solutions.

In agreement with the existing literature (see for example \cite{poor}), we found that generally the BSSN formalism is less able to handle gauge waves compared to the ADM
\footnote{This difference is mainly due to truncation and roundoff error propagation. The BSSN formalism involves a higher number of products than the ADM. Just as an example, by using double precision (that we use throughout the paper), even the initial Hamiltonian is already $\sim 10^{-16}$. We checked that by switching to quadruple precision the errors reduces to $H\sim 10^{-32}$.}. The present results are also in agreement with the simulations in \cite{campan,daverio}, although the Hamiltonian error shown here is at least one order of magnitude smaller than the one presented in the above works. This good performance is evidently due to the precision of the spectral representation, accompanied by the RSC and the dealiasing filters.

\subsection{The robust stability test}
The robust stability test is a classical initial condition for numerical relativity, particularly stressful from the numerical point of view, that makes use of random, small perturbations in the initial data. This testbed is able to initiate numerical instabilities that might exponentially grow in time, arising from small scales and propagating back to larger structures. These growing modes are quite dangerous since they might be confused with physical mechanisms. It is therefore important to inspect these numerical instabilities in order to prevent numerical artifacts. 
\begin{figure}
\hspace{-10pt}
\includegraphics[height=110mm,width=155mm]{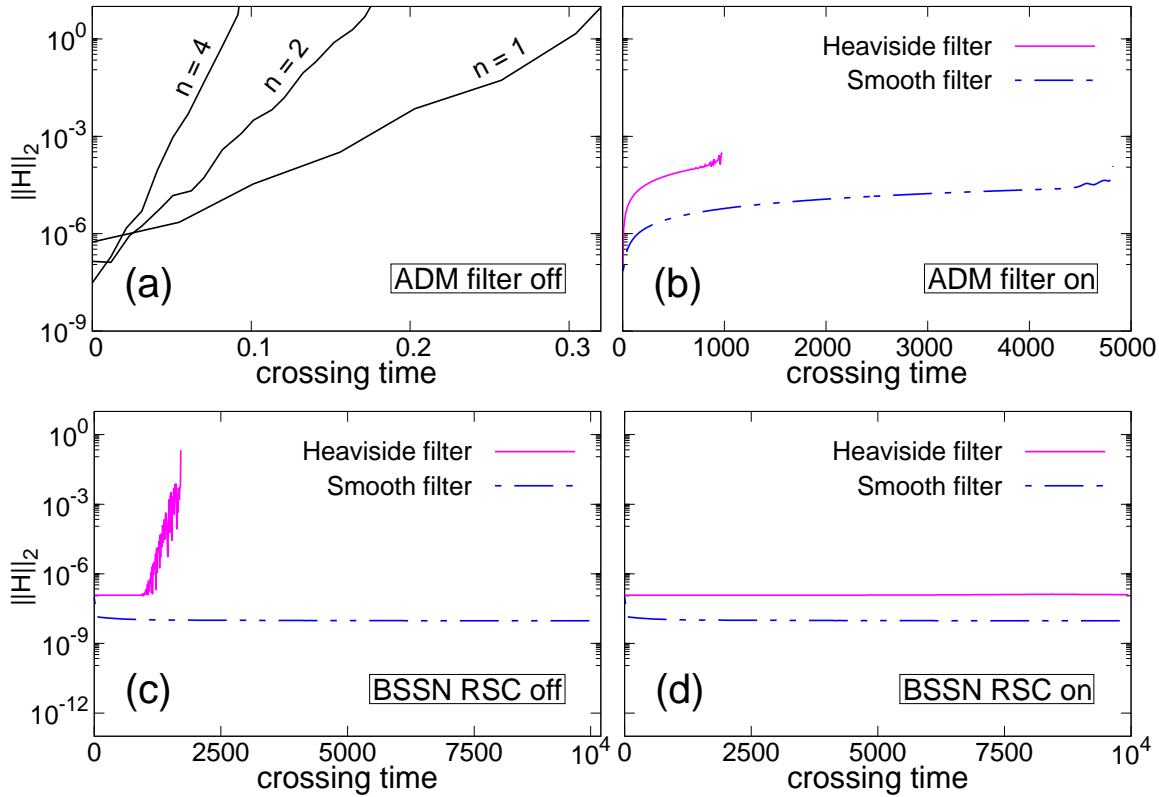}
\caption[ ]{\footnotesize{}Hamiltonian constraints $||H||_2$ for the robust stability test. (a) Results for the ADM code, with $n=1$ (RUN$_9$),  $n=2$ (RUN$_{10}$) and  $n=4$ (RUN$_{11}$). No RSC and no anti-aliasing filters have been used. (b) Same ADM simulations, with filters at $k^*=N/4$, as described by equations (\ref{filteralias2})-(\ref{filteralias1}), $\Delta t= 10^{-3}$ (fixed, no RSC) and $n=1$. These runs are labelled as RUN$_{12}$ and RUN$_{13}$. (c) Results from the BSSN code, using filters and no RSC with $\Delta t = 10^{-3}$ (RUN$_{14}$ and RUN$_{15}$). (d) Same simulations, for the BSSN, with the RSC (RUN$_{16}$ and RUN$_{17}$). Numerical stability has been achieved with the combined use of filters and time-step conditioning.}
\label{fig:31}
\end{figure}

The initial configuration consists of small random perturbations to the  Minkowski space, where $ds^2 = -dt^2 + dx^2 + dy^2 + dz^2$, with the unperturbed metric $\eta_{ij}$. The numerical noise is distributed among every field. Since we base our numerical algorithm on the use of FFT's, we vary several resolutions, keeping the number of meshes as a power of $2$. In order to follow the same configuration of Ref.~\cite{Alc}, we chose a spatial domain $x, y, z \in [0,1]$, a number of grid points $N_x =50 \rho_n$, where the mesh size is therefore $\Delta x= 1/(50 \rho_n$) and the initial time-step is $\Delta t=0.01 / \rho_n$. Here we vary $\rho_n = 64n/50$, with $n = 1,2,4$. To speed up calculations, we use only four grid points in the $y$ and $z$ directions. We solve the equations by using the harmonic gauge, i.e. $f(\alpha)=1$ in the slicing equation (\ref{slicing}). In the ADM tests, the initial conditions are given by
\begin{equation}
    \gamma_{ij}=\eta_{ij} +\varepsilon_{\gamma_{ij}}~,~~~K_{ij}=\varepsilon_{K_{ij}},~~~\alpha=1+\varepsilon_{\alpha},
\end{equation}
where $\varepsilon_f$ are random independent perturbations with 
\begin{equation}
\label{epsi}
    \varepsilon_f \in \left[-{10^{-10}}/\rho_n^2,{10^{-10}}/\rho_n^2\right].
\end{equation}	
In our robust stability tests we vary the resolution ($n$), the type of the dealising filter (Heavyside or smooth), and we vary from fixed time-step to the RSC. The goodness of our numerics has been monitored by looking at the evolution of the constraints in time, such as the $L_2$-norm of the Hamiltonian constraint.

In figure \ref{fig:31} (a) we report the Hamiltonian error, for different resolutions, namely for $n=1,2,4$. No anti-aliasing filters have been used for the runs in this panel. As it can be seen, the code becomes suddenly unstable. These runs are reported in table~\ref{table}, labeled as RUN$_9$--RUN$_{11}$. In order to show the benefit of the anti-aliasing filters and the robustness of the code, a second set of tests has been performed, by using filters at $k^*=N/4$, but without RSC. A time-step of $\Delta t=10^{-3}$ and $n=1$ has been chosen. The $L_2$-norms of the Hamiltonian constraint are shown for these cases in figure~\ref{fig:31} (b). The simulations show that, with the ``Heaviside'' filter in (\ref{filteralias2}), the code becomes now stable until $t\sim 1000$, while is even more stable using the smooth filter. These runs are reported in table~\ref{table}, as RUN$_{12}$ and RUN$_{13}$.

We complete the robust stability test campaign with the BSSN approach and finally introduce the RSC condition. The parameters are the same of the ADM previous tests. For the slicing condition we used the harmonic and hyperbolic Gamma-driver slicing in equations (\ref{slic2}) and (\ref{slic3}). The tests are carried out for $t=10^4$ crossing times, at most. First we compare the two filters described by equations (\ref{filteralias2})-(\ref{filteralias1}), but keeping a fixed time-step of integration. In the bottom left panel of figure \ref{fig:31}, we show the $||H||_2$ for the two filters. Note that using the Heaviside filter the code crashes at $t\sim 2000$ crossing times (RUN$_{14}$), while using the smooth version it remains more stable (RUN$_{15}$). In panel (d), finally, we show the same tests with the RSC, by using an adaptive time refinement with $C=1/4$, as described in section \ref{rsc1}. In this case, the full algorithm stabilizes the simulation until $t=10^4$ crossing times (RUN$_{16}$--RUN$_{17}$). Stability of the run has been achieved with the combined use of filters and adaptive time-integration.

\subsection{New initial data: the standing waves}
Here we propose a very simple, 1D, initial set of data that satisfies equations (\ref{constraints})-(\ref{Hadm}) and that can be used to test codes in numerical relativity, with various amplitudes.  This initial condition consists of a simple initial data for the extrinsic curvature only, on an initial flat metric, that suddenly produces some non-propagating, oscillatory modes. The physical pattern consists of a ``bouncing metric''. This solution to the constraint conditions is not supposed to have an immediate physical application, as in many of the other testbeds, but only for testing purposes. However, it might have some applications and similarities with cosmological models of inflation. For this regard, see for example Adamek et al. \cite{adamek}.
\begin{figure} [t]
\hspace{-14pt}
\includegraphics[height=110mm,width=155mm]{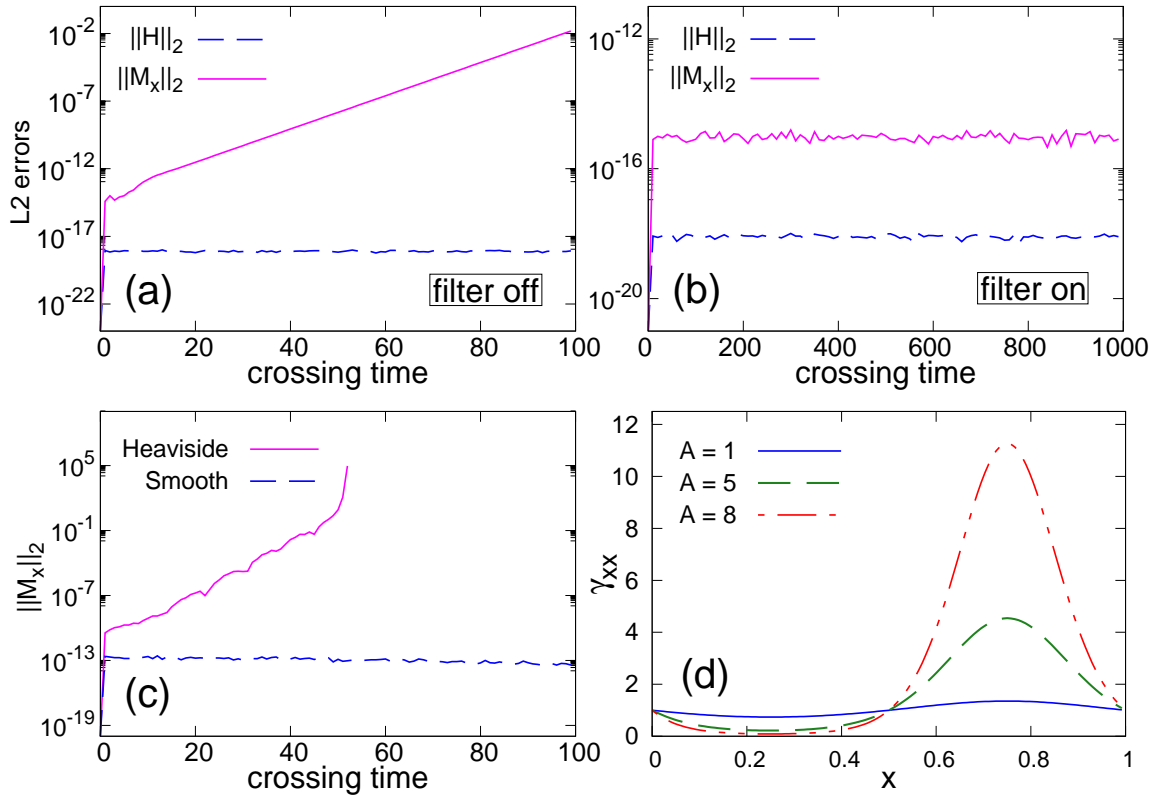}
\caption[]{\footnotesize{} (a) $L_2$ norm of Hamiltonian and $x-$momentum violation for the standing waves (RUN$_{18}$). Without an anti-aliasing filter, the Hamiltonian constraint remains constant while the $x-$momentum constraint grows exponentially. The evolution is carried out for $t=100$ crossing times. (b) Same simulations but with a smooth filter at $k^* = N/3$: both constraints remain negligible (RUN$_{19}$).  (c) The $x-$momentum constraint for the nonlinear test ($A=8$), using both anti-aliasing filters at $k^* = N_x/3$ (RUN$_{20}$ and RUN$_{21}$). The comparison emphasizes again the goodness of the smooth filter. (d) $\gamma_{xx}$ for the standing wave test, for different amplitudes. As the amplitude increases, the wave becomes more asymmetric.}
\label{hi100}
\end{figure}

The initial spatial metric $\gamma_{ij}$ obeys Minkowski flat space, which means zero curvature and where Ricci tensor vanishes ($R=0$). In this case, the  Hamiltonian and the momenta constraints reduce to
\begin{eqnarray} 
\label{kkk}
    K^2-K_{ij} K^{ij}=0,\\
    D_i(K^{ij}- \gamma^{ij} K) =0,
\end{eqnarray}
respectively. We now chose an initial perturbation only for the extrinsic curvature $K_{ij}$, with a Minkowski flat initial metric. With this choice, one has:
\begin{eqnarray}
\nonumber
    \gamma_{ij}= \eta_{ij}
    ~,~~~~~	K_{xx}= A~\sin(2 \pi x),
\end{eqnarray}
and $K_{ij}=0$ otherwise, which satisfies the set of constraints in equation (\ref{kkk}). Unlike the previous case of gauge waves, this condition does not lead to a wave propagation in space but to standing waves of the metric. One can stress the code by increasing the amplitude of the extrinsic curvature perturbation $A$, as follows.

In figure \ref{hi100} we report the time evolution of the metric, for several ADM tests. For RUN$_{18}$, a small amplitude $A=10^{-1}$ and no anti-aliasing filter has been used. We chose a spatial domain  with $x \in [ 0, 1 ]$ and a spatial grid $N_x = 64$. The time-step is $\Delta t = 10^{-3}$ and the harmonic slicing [$f(\alpha)=1$] has been chosen. The simulation has been carried out until 100 crossing times. We have carried out a second test (RUN$_{19}$) with same parameters but choosing a smooth filter with $k^* = N_x/3$, as reported in figure \ref{hi100} (b). It is evident that the filter improves the stability, in fact this run is carried out until 1000 crossing times. The top panels of figure \ref{hi100} show that the Hamiltonian constraint remain constant in both cases, but the $x-$momentum constraint (for the symmetry of the problem the $y-$momentum and $z$-momentum are zero) grows without filter and remains constant with $k^* = N/3$.

In order to test a more nonlinear case, a second, high-amplitude condition has been used, where we chose $A=8$, a spatial grid with $N_x=128$ points and a $k^* = N/3$ filter. All the other parameters have been chosen as in the previous tests, as summarized in table \ref{table}. The higher amplitude of $A$ now induces asymmetry in the standing wave. The metric starts to bounce and becomes asymmetric. Again, this test has been performed twice: a first test  (RUN$_{20}$) using the standard truncated filter described by equation (\ref{filteralias2}) and a second one  (RUN$_{21}$) with the filter in equation (\ref{filteralias1}).  The time-evolution of $L_2$ norm of $x-$momentum constraint, computed using the definition in equation (\ref{norml2}), is shown in the panel (c) of figure \ref{hi100}, for both cases. It is clear that the smooth filter works better than the Heaviside filter.

We ran the same tests for the BSSN code (not shown), setting $A=10^{-1}$, a spatial grid with $N_x = 64$ points, a time-step $\Delta t = 10^{-4}$ and the ``1 + log'' slicing (i.e. $f(\alpha) = 2/\alpha$ in equation (\ref{slicing}) and $\beta^i=\partial_t \beta^i=0$). The tests are carried out until 100 crossing times. Similarly as in the previous ADM tests, the smooth filter better stabilizes the code.

\subsection{The gravitational wave test}
\label{GW}
In this subsection we test our two codes, the ADM and the BSSN spectral algorithms, via linear gravitational waves. In particular we check whether they are able to retain accuracy, for long spacetime travels, by comparing amplitudes and phases with the well-known analytical solution.  This solution to the linearized Einstein field equation can be written as
\begin{equation}
    ds^2 = - dt^2 + dx^2 + (1+b)\,dy^2 + (1-b)\,dz^2,
\label{grav}
\end{equation}
where $b = A~ \sin \left[ \frac{2 \pi (x-t)}{d}\right]$ and $d=1$ is the linear size of the propagation domain. The wave amplitude is small, in order to satisfy the linear regime conditions. The metric in equation (\ref{grav}) represents the transverse-traceless gauge in which the wave amplitudes are purely spatial, with null trace and transverse to the propagation direction. It describes a wave propagating along the $x$ axis, with the polarization aligned with the $y$ and $z$ axes \cite{onda}. This metric is written in Gauss coordinates, that is $\alpha = 1$ and $\beta^i = 0$. Using this metric, the extrinsic curvature reduces to $K_{ij}=-\partial_t \gamma_{ij}/(2\alpha)$, so that the nontrivial components reduce to $K_{yy}=\frac{1}{2} \partial_t b$ and $K_{zz}=-\frac{1}{2} \partial_t b$. Since this wave propagates along the $x$-axis and all derivatives are zero in the $y$ and $z$ directions, the problem is essentially 1D.
\begin{figure}
\hspace{-12pt}
\includegraphics[height=118mm,width=164mm]{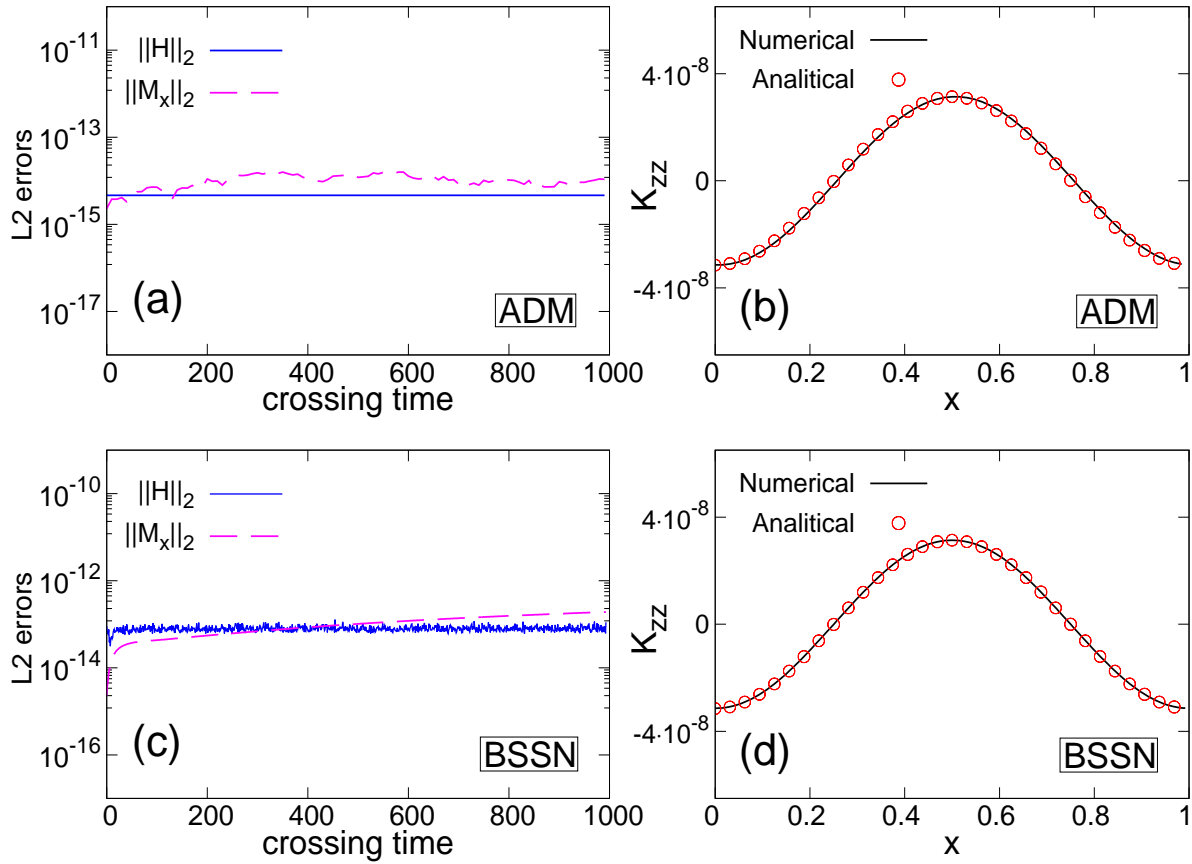}
\caption[ ]{\footnotesize{}Gravitational wave test. (a) $L_2$ norm of ADM constraints (RUN$_{22}$). A smooth filter with $k^* = N_x/4$ has been used. (b) Comparison of $K_{zz}$ and its exact solution, at $t=1000$ for the same run. (c) Constraints for the BSSN simulation (RUN$_{23}$). (d) Comparison of the wave form of $K_{zz}$  with the exact solution at $t = 1000$, for the same run. A smooth filter with $k^* = N_x/8$ has been used for this simulation.}
\label{fig:39}
\end{figure}
We chose a spatial domain $x \in [0,1]$ on a spatial grid with $N_x = 64$, with an initial $\Delta t=10^{-3}$ and the harmonic slicing. We use a very small amplitude, namely $A=10^{-8}$, in order to satisfy the linearity of the evolution \cite{Dum}. The evolution is carried out up to $t = 1000$ and a smooth filter, described by equation (\ref{filteralias1}), with $k^*=N/4$, has been used.

Figure \ref{fig:39} (a) represents the time evolution of the $L_2$ norm of ADM constraints. The errors are well bounded. In the panel (b) we compare the wave-form of $K_{zz}$ at the final time, i.e. at $t=1000$, showing overall an excellent agreement between numerical and exact solutions. This run is reported in table~\ref{table} as RUN$_{22}$.  We performed an analogous test with the BSSN formalism, choosing $N_x = 128$, $\Delta t=5 \times 10^{-5}$, with the harmonic slicing and the ``Gamma-driver'' shift condition. The evolution is carried out again for a time of $t = 1000$ and the smooth filter with $k^*=N_x/8$ has been used. This run is labeled in table~\ref{table} as RUN$_{23}$. The bottom left panel of figure \ref{fig:39} shows the $L_2$ Norm of the Hamiltonian, and the bottom right panel of the same figure shows the excellent agreement with the analytical solution for the waveform of $K_{zz}$, at the final time. It is important to stress that, in the BSSN runs, the choice of a stronger dealiasing filter has been used in order to obtain a longer time, stable and accurate simulation. We also point out that here we used a finer grid, so that the simulation corresponds essentially to the ADM runs. More points are now needed since the BSSN is less accurate. As discussed in previous sections, this is related to the fact that BSSN is a more elaborated formalism, where a larger number of products leads to higher truncation errors and therefore to less precise solutions \cite{Alc}.

In summary, the gravitational wave test therefore is well solved via both algorithms, the smooth filter stabilizes the code for very long times, giving a very good representation of the solution.

\subsection{The Gowdy spacetime}
\begin{figure}[htbp]
\hspace{-10pt}
\includegraphics[height=58mm,width=160mm]{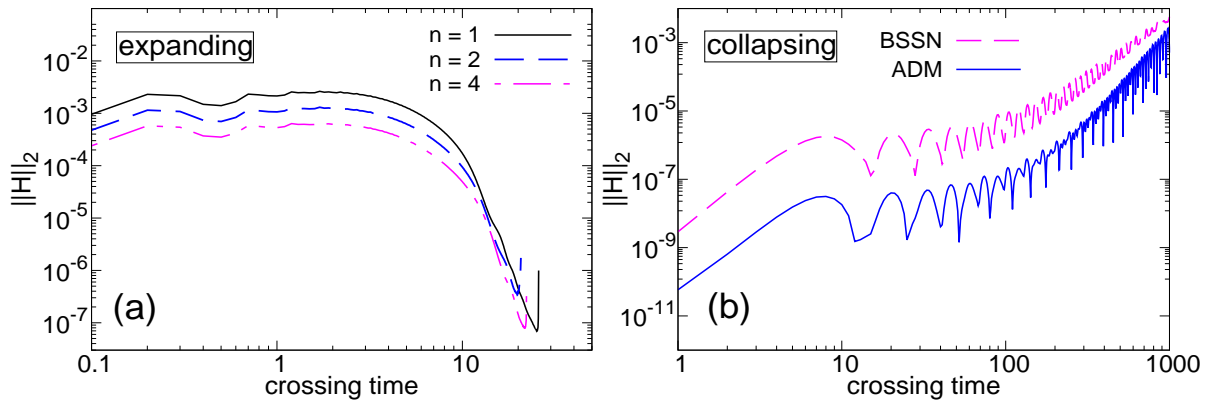}
\caption[ ]{\footnotesize{} (a) Constraint violation $||H||_2$ for the BSSN pseudo-spectral code, for the expanding Gowdy spacetime, at different resolutions (RUN$_{24}$--RUN$_{26}$). (b) Same, for both ADM (RUN$_{27}$) and BSSN (RUN$_{28}$), for the collapsing spacetime and $n=1$.}
\label{gowdy1}
\end{figure}
All the tests described so far considered perturbations to the flat metric. In this section we deal with a genuinely curved exact solution, known as the polarized Gowdy spacetime \cite{daverio,campan}.  The Gowdy waves are vacuum cosmological models used to test codes in a strong field context and present a serious challenge for numerical relativity. These particular solutions of the vacuum Einstein equations on the 3-torus describe an expanding/collapsing universe, containing plane polarized gravitational waves where
\begin{equation}
    ds^2 = -\frac{e^{\lambda /2}}{\sqrt{t}} dt^2 + \frac{e^{\lambda /2}}{\sqrt{t}} dz^2 + t e^P dx^2 + t e^{-P} dy^2. 
    \label{gowdym}
\end{equation}
Here $\lambda$ and $P$ are functions of $z$ and $t$ only. The time coordinate $t$ is chosen such that time increases as the universe expands. The metric is singular at $t=0$ which corresponds to the cosmological singularity. We will carry out our tests in both time directions, i.e. in the collapsing and expanding dynamics. 

With the metric described by (\ref{gowdym}), the Einstein equations lead to \cite{Alc, gowdy}
\begin{equation} 
   P(z,t) = J_0( 2 \pi t) \, \cos(2 \pi z),  \label{p} 
\end{equation}
\begin{eqnarray} 
  \fl  ~~~~~\lambda(z,t) = -2\pi J_0(2 \pi t) \, J_1(2 \pi t)\,  cos^2(2 \pi z) + 2 \pi^2 t^2 \Big[ J_0^2(2 \pi t) + J_1^2(2 \pi t) \Big] \nonumber
    \\
    - \frac{1}{2} \Big[ 4 \pi^2 \Big[  J_0^2(2 \pi ) + J_1^2(2 \pi )  \Big] - 2 \pi  J_0(2 \pi) + J_1(2 \pi ) \Big],    \label{la}
\end{eqnarray}
where $J_0$ and $J_1$ are Bessel functions. Equations (\ref{p}) and (\ref{la}) yields a diagonal spatial metric of the type
\begin{eqnarray}
    \nonumber
    \gamma_{xx} = t e^P~,~~~\gamma_{yy} = t e^{-P}~,~~~\gamma_{zz} = \frac{e^{\lambda/2}}{\sqrt{t}}.
\end{eqnarray}
To complete the initial data conditions, the extrinsic curvature is obtained as a time derivative of the metric. Since the metric (\ref{gowdym}) depends only on the $z-$direction, the spatial evolution is 1D. We chose a spatial domain $z \in [ 0, 1 ]$, with $N_z =50 \rho_n$,  $\Delta z=1/(50 \rho_n)$ (see before), and a $\Delta t=5 \times 10^{-4} / \rho_n$. In all the Gowdy tests, we used a $k^* = N_z/5$, smooth, anti-aliasing filter.

In the expanding form, the time coordinate $t$ increases during the evolution of the simulation, $P$ approaches zero asymptotically and $\lambda$ increases linearly. Since the metric is singular at $t=0$, we set the initial data from the exact solution at $t=1$ and proceed forward in time.  Due to the exponential growth, such evolution may crash rather soon but it will test the accuracy of a code in a rather harsh situation. In figure \ref{gowdy1} (a) we report the $L_2$ norm of the Hamiltonian, for the three resolutions $n=1, 2, 4$ (RUN$_{24}$--RUN$_{26}$). The errors are bounded in all the tests and our results are in good agreement with the basic literature \cite{gowdy1, gowdy2}. Similar results have been found for the ADM simulation.

In the collapsing version, the time coordinate in the Gowdy metric can be transformed so that the initial singularity is approached asymptotically, backwards. The new time coordinate $\tau$ is defined by $\tau = \frac{1}{c}\, \ln(t/k)$. In other terms, we replace $t \rightarrow k e^{c\tau}$ and the time-step $\Delta \tau$ is chosen to be negative. We choose a particular value of $t_0$ such that the initial slice is far from the cosmological singularity and, following the standard approach \cite{Alc}, we chose the twentieth zero of the Bessel function, so that $J_0(2 \pi t_0) = 0$. Finally we get $c \sim 2\times 10^{-3}$ and  $k \sim 9.6\times 10^{13}$. In the panel (b) of figure \ref{gowdy1}, we present a comparison between the ADM (RUN$_{27}$) and BSSN (RUN$_{28}$) runs, for the collapsing direction. The ADM code shows a smaller error on the Hamiltonian error, but both schemes proceed until 1000 crossing times ($\tau = -1000$). This evolution is less challenging than the expanding case, since the lapse function is essentially an exponential in $\tau$ so that the spacetime is becoming less dynamical. On the other hand, the value of the physical metric $\gamma_{zz}$ at $t=t_0$ is of the order of $10^3$. Overall, as it can be seen, also in this more challenging test, the pseudo-spectral code is able to handle the numerical evolution, with violation errors that are comparable (or smaller) than in the existing literature \cite{Alc}.

\subsection{The Schwarzschild black hole}
\label{schws}
In this classical test we consider the stability of an isolated, Schwarzschild black hole (BH), in a fully 3D spatial geometry. The Schwarzschild metric is used as initial data to test the ability of the code to evolve BH spacetimes within the puncture approach \cite{camp, bh0, bh1, bh2}. The metric in isotropic coordinates can be written as \cite{book, Alc}

\begin{figure}[htbp]
\hspace{-4pt}
\includegraphics[height=62mm,width=157mm]{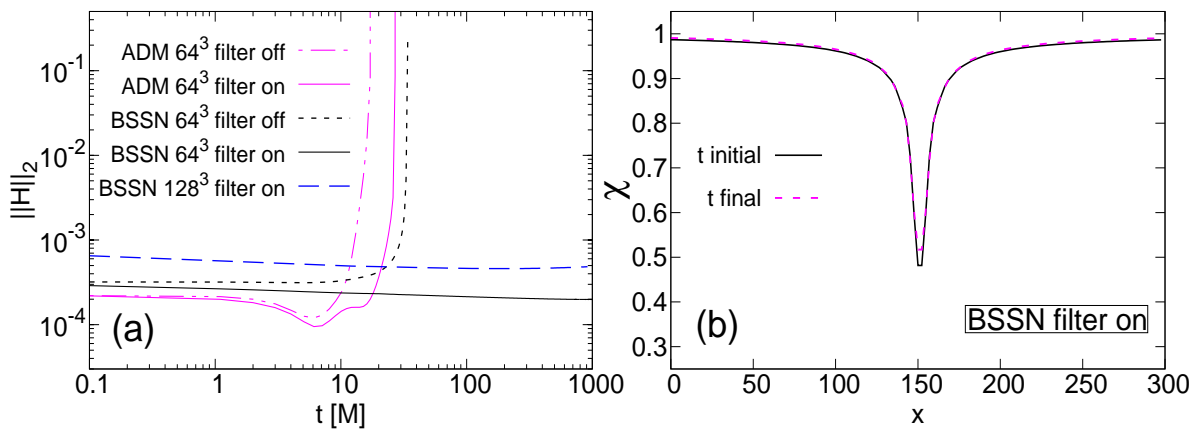}
\caption[ ]{\footnotesize{} Simulations of the Schwarzschild black hole. (a) Comparison of the time history of the error $||H||_2$, between the ADM and the BSSN codes (RUN$_{29}-$RUN$_{32}$). Note the log scale of the $x-$axis.
\reply{
    (b) Conformal factor $\chi = \psi^{-4}$, at different times of the BSSN simulation, with the same parameters as RUN$_{32}$, but higher resolution $128^3$. A smooth filter $k^* = N/2 $ has been used.
}
}
\label{bh}
\end{figure}

\begin{eqnarray}
    \nonumber
    ds^2 = \left( \frac{1-m/(2r)}{1+m/(2r)} \right)^2 dt^2 + \psi^4 \left( dr^2 + r^2 d\Omega^2 \right), 
\end{eqnarray}
where $r$ is the isotropic radius, $m$ is the mass of the black hole and
\begin{eqnarray}
    \nonumber
    \psi= 1 + \frac{m}{2r}
\end{eqnarray}
is the conformal factor. The initial spatial metric and the extrinsic curvature become
\begin{eqnarray}
    \nonumber
    \widetilde{\gamma}_{ij} = \psi^{-4} \gamma_{ij}= \chi ~ \gamma_{ij} ,~~~~~K_{ij}=0,
\end{eqnarray}
respectively. The lapse is initially set to one and the shift is vanishing everywhere. For a Schwarzschild black hole both the linear momentum and the dimensionless spin are set to zero \cite{york}.

We set the puncture mass $m=1.0M$ and the runs are carried out until $t=1000M$, corresponding to 1000 crossing times, unless the code becomes unstable. The spatial domain is $x, y, z \in [ 0, 300 ]$, with an isotropic spatial grid $N = 64^3$ and a time-step $\Delta t = 10^{-1}$. The 1 + log gauge condition is now used [i.e. $f(\alpha) = 2/\alpha$ in equation (\ref{slicing}) and $\beta^i=\partial_t \beta^i=0$], for both the ADM and BSSN tests.  
\reply{
    In order to handle Schwarzschild equilibrium, since the puncture limits differentiability of the solution, we shifted the position of the singularity by half grid-step, namely $\Delta/2$, for each Cartesian direction. In this regard, it is worth saying that exact spectral projections always allow to obtain the field in the continuum via a simple phase-shift of the Fourier coefficients \cite{YeungPope88}.
}

In the panel (a) of figure \ref{bh} we report the evolution of the Hamiltonian error $||H||_2$, for the ADM (RUN$_{29}$ and RUN$_{30}$) and the BSSN tests (RUN$_{31}$ and RUN$_{32}$), with and without the smooth dealiasing filter. Here we found that $k^* = N/2$ is enough to stabilize the simulation.  As expected, and as well known in the literature \cite{book, Alcu}, the ADM formalism is not well suited for the simulations of singular objects such as BH's. Even if the ADM ``beats'' the BSSN formalism in several numerical challenges (see previous sections) regarding the precision, the BSSN remains one of the most stable numerical approaches for massive stars dynamics. 
\reply{
    In figure \ref{bh} (b) we report the conformal factor $\chi = \psi^{-4}$ for a simulation with the same parameters of RUN$_{32}$ but with higher resolution ($128^3$). We report a 1D cut through the center of the domain. The simulation shows slightly higher non-conservation (probably due to the choice of the time-step), but it is extremely more precise, as we shall discuss later in the convergence tests of the Appendix. For this version of the algorithm, namely the filtered BSSN, the simulation is stable (due to the constraint propagation properties of BSSN), accurate and convergent.
}

As is clear from figure \ref{bh}, the BSSN is able to simulate black holes dynamics, accurately, for long times. Even with very small dealising filters, with $k^*=N/2$, the BSSN retains the black hole shape and keeps a very small error on the constraints. However, the small violation of the periodicity, together with the high accuracy of the spectral code, might produce wiggles that propagate from the discontinuous boundary. In the next section, we will deal with these effects and cure this pathology by introducing a new numerical strategy, where we still keep the high precision of the pseudo-spectral strategy.

\section{The implicit hyperviscous boundary method}
\label{secihb}

We study black hole dynamics via a head-on collision of two massive objects. We propose a new method in order to treat the boundary conditions with our pseudo-spectral, periodic code, based on the use of absorbing hyperviscous boundaries.  As demonstrated in previous sections, our pseudo-spectral technique, together with an appropriate choice of the dealiasing and the time-step, is able to handle all the classical numerical testbeds. The main advantage of the pseudo-spectral code is of being relatively fast (since it stems on FFTs) and incredibly precise (since it relies on an exact decomposition). It would be hard to beat the precision of such methods with finite difference or finite volume/elements techniques, at a given resolution \cite{orszag}. On the other hand, it is evident that the main weakness of the method relies on the boundary conditions that need to be periodic for this particular spectral decomposition. Periodic boundaries are not always a good choice for general relativity problems, as for example in the case of compact object mergers. Here we present an additional algorithm that helps to overcome this limitation, which is based on the use of dissipating boundaries, combined with implicit time integration. This Implicit Hyperviscous Boundary method (IHB) has been tested with gravitational waves and compact objects dynamics. The method is relatively easy-to-implement and efficient.

The following IHB method has been inspired by works in hydrodynamics and magnetohydrodynamics, with applications to turbulence and to stellar dynamo. It is indeed possible to use Cartesian geometries in order to simulate spherical domains, using for example volume penalization methods \cite{Schneider, Dobler}, or by using hyperviscous forces \cite{Servidio}, starting from a given distance from the center of the Cartesian lattice. In this way it is possible, for example, to embed a spherical domain inside a cube, by dissipating the regions close to the corners, with a radial envelope. Here we re-elaborate this idea and apply it to numerical relativity, in a pretty straightforward way.

Hyperviscoscous terms of the type $\nu_n \nabla^n f$, with $\nu_n$ a numerical coefficient, are able to dissipate quickly ripples and numerical artifacts, but they might reduce dramatically time-step since they involve high-order derivatives and fast diffusion. For the method presented here show results with fourth-order dissipation (but the technique can be extended to any order.) We make use of a technique for the hyperviscous terms based on the implicit, second-order Crank-Nicholson method \cite{CNbook}. Hereafter we will apply this technique only to the BSSN code, which is a stable method for solving problems such as the binary mergers and black holes dynamics.

For any variable of the BSSN method, we advance in time the solution $f^{\{ideal\}}({\bf x}, t)$ by using the second order Runge-Kutta described in equation (\ref{rk2}). Simultaneously, we evolve in time an accompanying, twin field $f^{\{H\}}({\bf x}, t)$ which obeys to the same BSSN equation but is also subject to hyperviscous dissipation, being therefore highly damped. We advance in time this hyperviscous field $f^{\{H\}}({\bf x}, t)$ by using a Crank--Nicolson, semi-implicit method (CN). The CN integration can be easily matched with the Runge--Kutta method and it is numerically very stable with respect to viscous dynamics. For any BSSN hyper-dissipated variable we have an evolution equation 
\begin{equation}
   \frac{\partial f^{\{H\}}({\bf x}, t)}{\partial t} = N({\bf x}, t) - \nu_4 \nabla^4 f^{\{H\}}({\bf x}, t), 
   \label{Hyp}
\end{equation}
where $N$ indicates the BSSN evolution equation, $\nu_4$ is the hyper-viscous coefficient. In the case of the ``ideal'' BSSN set of equation, $\nu_4=0$ and taking the Fourier transform, from the typical Runge-Kutta 2-steps method one gets
\begin{eqnarray}
\nonumber
    \frac{\widetilde{f}^{\{ideal\}}_{{\bf k}} \left(t^* + \Delta t \right) - \widetilde{f}^{\{ideal\}}_{{\bf k}} (t^*)}{  \Delta t} =  \widetilde{N}_{{\bf k}}\left(t^* + \frac{\Delta t}{2} \right),
\end{eqnarray}
at a given time $t^*$, for any Fourier coefficient $\widetilde{f}^{\{ideal\}}_{{\bf k}}$,  at a given wavenumber $\bf k$.  At each time $t^*$ we start with $f^{\{H\}}({\bf x}, t)=f^{\{ideal\}}({\bf x}, t)$. Proceeding in a similar way for the hyper-viscous accompanying field $f^{\{H\}}$, and making  implicit the linear hyper-viscous term on the $rhs$, one obtains 
\begin{eqnarray}
\nonumber
  \fl  ~~~~~~~\frac{\widetilde{f}^{\{H\}}_{{\bf k}} \left(t^* + \Delta t \right) -
    \widetilde{f}^{\{H\}}_{{\bf k}} (t^*)}{  \Delta t} =  \widetilde{N}_{{\bf k}}\left(t^* +
    \frac{\Delta t}{2} \right)
    - \frac{\nu_4 k^4}{2} \left[ \widetilde{f}^{\{H\}}_{{\bf k}}
    \left(t^* + \Delta t \right) + \widetilde{f}^{\{H\}}_{{\bf k}} (t^*) \right].
\end{eqnarray}
At this point, for any damped field, it is easy to get the time-splitting, second order CN formula, 
\begin{eqnarray} 
   \fl  ~~~~~~~~~~~~~\widetilde{f}^{\{H\}}_{{\bf k}}\left(t^* + \Delta t \right)  =
    \left[ \frac{1- \frac{\nu_4 k^4}{2}\Delta t }{1+ \frac{\nu_4 k^4}{2}\Delta t } \right]
    \widetilde{f}^{\{H\}}_{{\bf k}} (t^*)
    +\left[ \frac{\Delta t}{1+ \frac{\nu_4 k^4}{2}\Delta t} \right] \widetilde{N}_{{\bf k}}\left(t^* + \frac{\Delta t}{2} \right), 
\label{cki}
\end{eqnarray}
which is stable and particularly easy to implement in pseudo-spectral codes. For all the BSSN variables, we advance in time equation (\ref{cki}) for both the ideal fields $\widetilde{f}^{\{ideal\}}_{{\bf k}}$ ($\nu_4=0$) and the hyperviscous  counterparts $\widetilde{f}^{\{H\}}_{{\bf k}}$ ($\nu_4\ne 0$). By taking the inverse transform now one has both fields $f^{\{ideal\}}({\bf x}, t+\Delta t)$ and $f^{\{H\}}({\bf x}, t+\Delta t)$ and then has to deal with the solutions match. 

The scope is to suppress spurious effects only at the boundaries, therefore we interpolate between the two solutions gradually (or in a Heaviside way) going towards the borders, only in a region close to the boundaries.  With this procedure we retain the pure, ideal dynamics of the BSSN formalism in the center of the box. Overall, the price to pay for the IHB is that we have to double the time of integration, slowing down the computation. On the other hand, the method gives clear benefits on the stability and the goodness of the solutions, as we shall see in the next subsections, where we present two tests.

\subsection{Gravitational wavepacket absorption}
\begin{figure}
\hspace{-18pt}
\includegraphics[height=95mm,width=165mm]{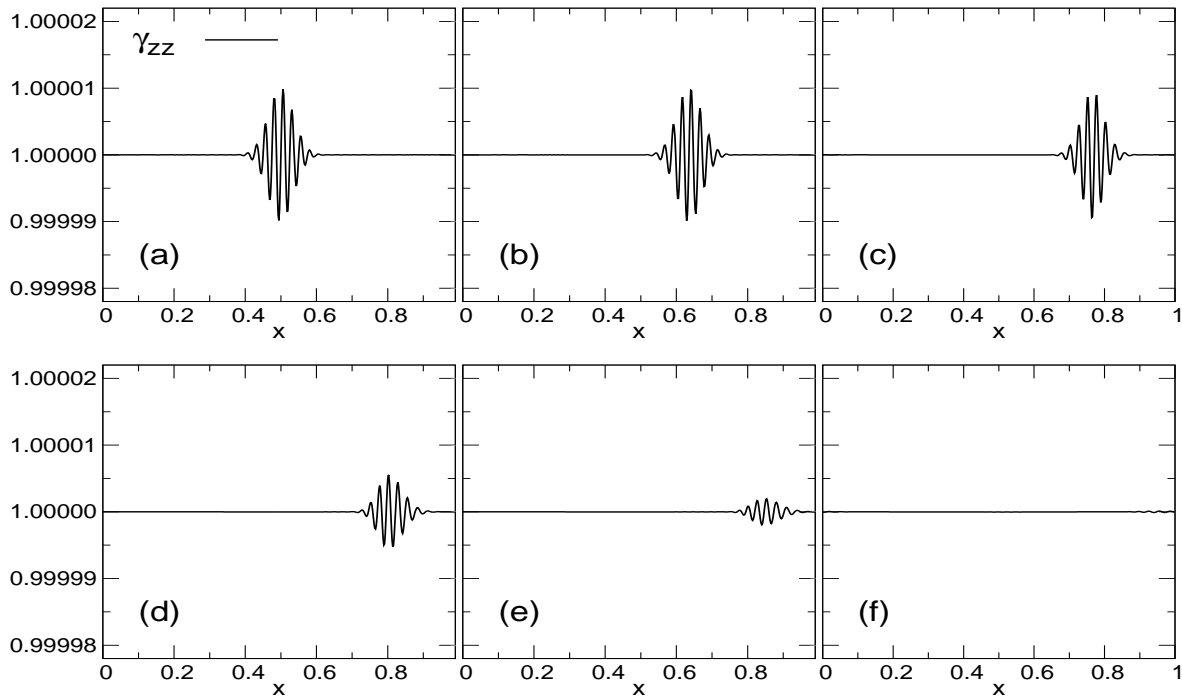}
\caption[Gravitational wavepacket absorption]
{\footnotesize{}Absorption of a 1D wavepacket of gravitational waves using our Implicit Hyperviscous Boundary (IHB) method, for the RUN$_{33}$.
The panels show the section of the physical metric $\gamma_{zz}$ at different times, i.e. $t=0,\, 0.14,\, 0.26,\, 0.32,\, 0.37,\, 0.48$.}
\label{gwpck}
\end{figure}

In order to test the IHB method, we check whether the algorithm is able to absorb fluctuations that move outward, from the center of the domain toward the boundaries. At the initial time, we build a gravitational wavepacket, namely a windowing of the gravitational wave at the center of the domain via a Gaussian-shaped filter, as shown in figure \ref{gwpck}-(a). As in the linearized gravitational wave test (see section \ref{GW}), the metric is given by equation (\ref{grav}). We initialize the perturbation $b(x,t=0)$ as a wavepacket that propagates along the $x-$axis, with a Gaussian window of the type $\exp[ -(\frac{x-x_0}{\sigma})^2 ]$. We set an amplitude of $A=10^{-5}$, the wavenumber $k=40$ and the width of the window $\sigma = 0.05$ is small enough in order to localize the waves in the center, namely $x_0 = L_0/2 = 1/2$, where $L_0=1$ is the domain.

In order to match the two solutions $f^{\{ideal\}}({\bf x}, t)$ and $f^{\{H\}}({\bf x}, t)$ in the IHB technique, we chose a linear matching when distance approaches to the borders. In particular, at the end of each time-step, we impose for the BSSN variables 
\begin{equation}
   \fl	f({\bf x}, t)= \cases{f^{\{ideal\}}({\bf x}, t)&for $ |{\bf x} - {\bf x}_0|\leq \lambda $, \\
	f^{\{ideal\}}({\bf x}, t)\left[2-\frac{|x-x_0|}{\lambda}\right]+f^{\{H\}}({\bf x}, t)\left[\frac{|x-x_0|}{\lambda}-1\right]& \{otherwise\}.}
	\label{matchh}
\end{equation}
In the above matching-condition we chose $\lambda=L_0/4$, namely one quarter of the box. In this test we set the hyperviscous coefficient $\nu_4 = 10^{-7}$. This test is labeled as RUN$_{33}$ in table \ref{table}.

With our IHB approach, when the packet is far from the boundaries, the BSSN evolution equation $f^{\{ideal\}}({\bf x}, t)$ dominates in equation (\ref{matchh}) -- all the field equations evolve in the standard way. Otherwise, when the waves approaches to the boundaries, the hyperviscous terms of the type $\nu_4 \nabla^4 f$ dominate, and the wavepacket is totally absorbed. The time history of the propagating gravitational wave is reported in figure \ref{gwpck}. As it can be seen, the modulation moves coherently from the center of the domain, toward $L_0=1$. When the wavepacket enter the region of the viscous boundary, at $x=\frac{3}{4}L_0$, it starts to ``feel'' the hyperviscous damping in equation (\ref{Hyp}). Before approaching to the borders, the wave is totally absorbed. Similar results (not shown here) have been obtained with other matching conditions, discussed in the next subsection. The method can be easily exported to 3D gravitational dynamics, in order to suppress boundary effects, as we will see in the next, final experiment.

\subsection{Head-on collision of black holes}
\label{eado}
In this last section we test the IHB technique via one of the most stressful conditions, namely the head-on collision of two black holes. In this test, the initial conditions consist of two non-rotating, massive compact objects. We test this black hole crash only via the most numerically stable approach, namely the  BSSN spectral code, with a smooth filter in equation (\ref{filteralias1}) and with the RSC active.

The parameters of the initial conditions are the following. The three-dimensional computational domain is $x, y, z \in [0,25]$ and flat Minkowski spacetime is imposed as a boundary condition everywhere. We chose two identical black holes with masses $m_1 = m_2 = 0.5M$, and $\bf{C_1} = $ $\{ 10.5,12.5,12.5 \}$ and $ \bf{C_2} =$ $ \{ 14.5,12.5,12.5 \}$, where the ${\bf C}_j$ are the locations of the two punctures. These parameters correspond to an initial separation of the BHs a little larger to that of an approximate innermost stable circular orbit (ISCO), as determined in \cite{43}. The two punctures consist of two Schwarzschild solutions with initially zero linear momentum and spin, as described in section \ref{schws}.

\begin{figure} [h]
\hspace{-10pt}
    \includegraphics[height=105mm,width=159mm]{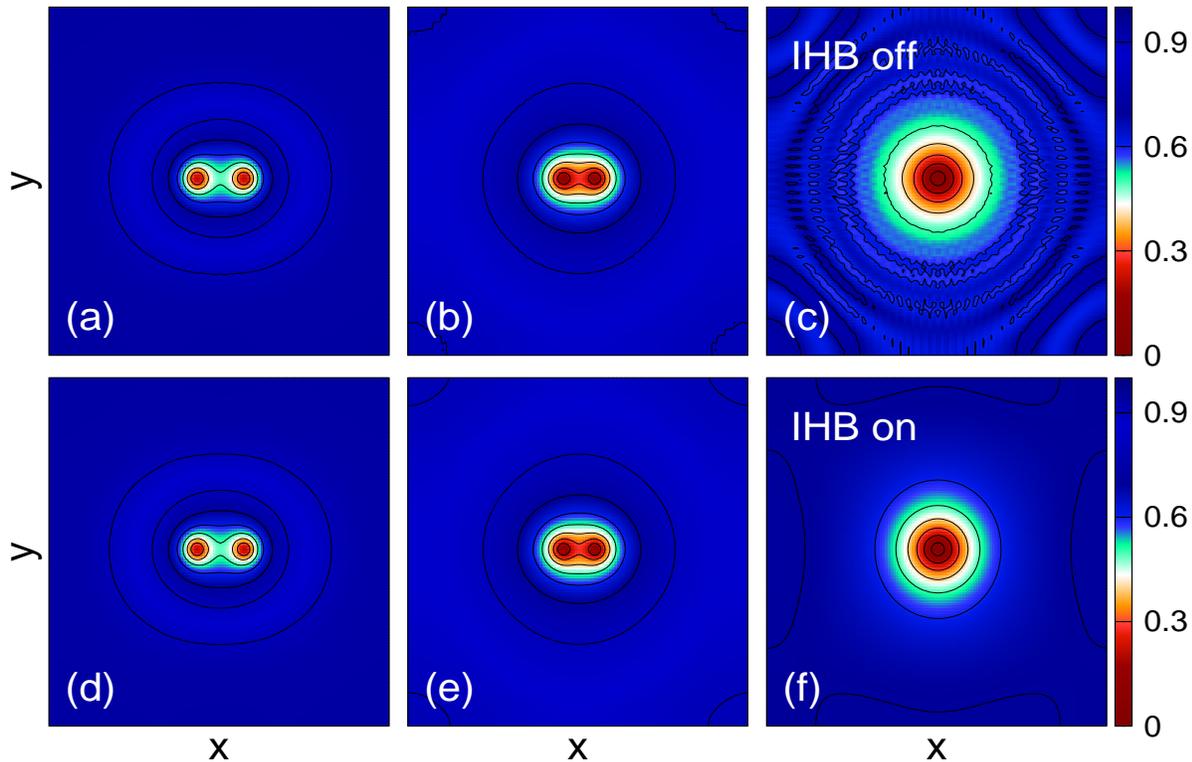}
    \caption{Time evolution of the contour surfaces of the lapse $\alpha$ for the head-on collision of two puncture black holes of equal mass $m_1 = m_2 = 0.5M$ at times $t = 1, 13$ and $25M$ (from left to right). Top row: without implicit hyperviscous boundaries (IBH) (RUN$_{34}$). Bottom row: with IHB (RUN$_{35}$). }
    \label{headon}
\end{figure}

As suggested by \cite{Dum}, the lapse is initially set to:
\begin{eqnarray}
\nonumber
    \alpha = \frac{1}{2} \left( \frac{1-\frac{m_1}{2 r_*^-}-\frac{m_2}{2 r_*^+}}{1+\frac{m_1}{2 r_*^-}+\frac{m_2}{2 r_*^+}}
    \right),
\end{eqnarray}
where $r_* \stackrel{d}{=} \left( r^4 + 10^{-24}\right)^{\frac{1}{4}}$   and $r$ is the coordinate distance of a grid point from the puncture.
\begin{figure} [h]
\centering
    \includegraphics[height=80mm,width=120mm]{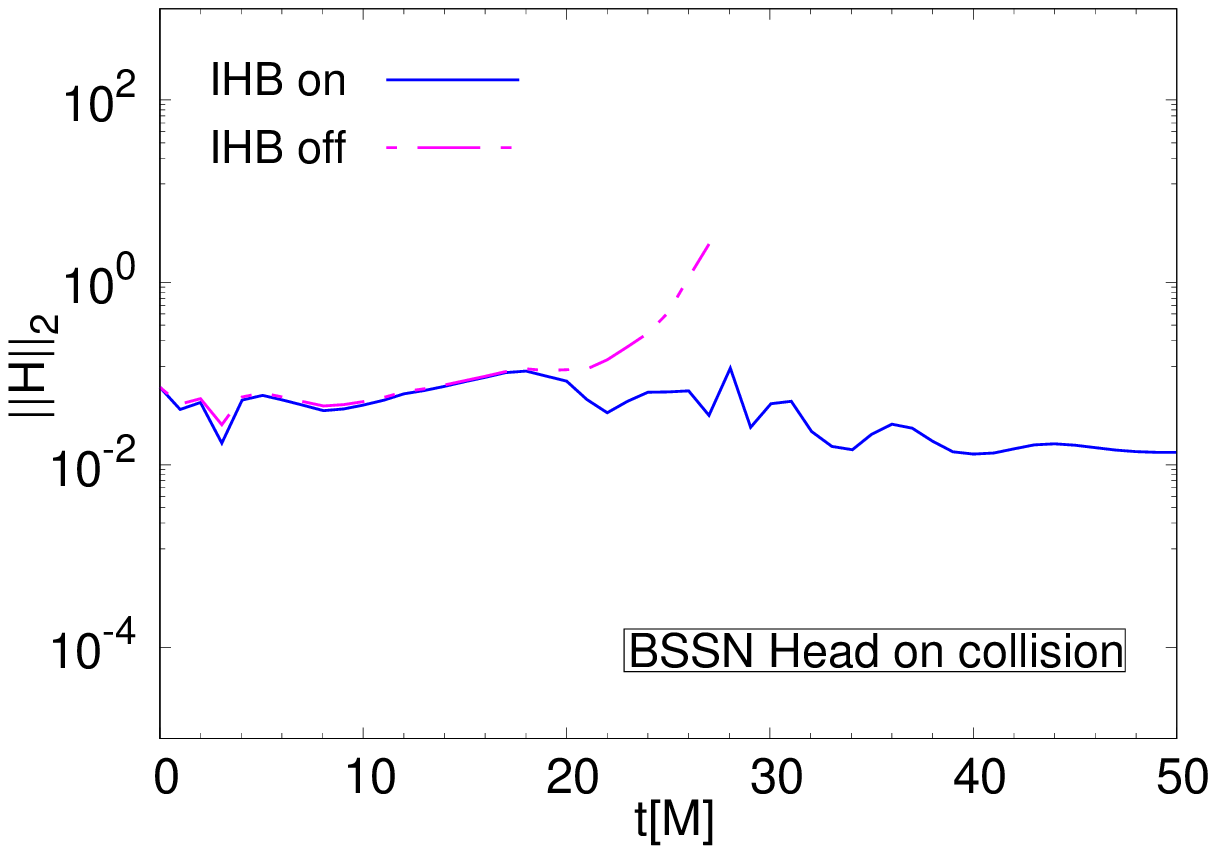}
    \caption{L2 errors of the Hamiltonian constraint for the BH's collision. The dash-dotted line represents the violation for the simulation without IHB (RUN$_{34}$). The code crashes at $t \sim 30M$. The solid line represents the same test with IHB (RUN$_{35}$). The Hamiltonian is well bounded and the simulation is carried out until $t = 50 M$.}
    \label{h2headon}
\end{figure}
We present two types of runs for the head on collision. We evolved a first simulation (RUN$_{34}$) without the IHB method. We carried out the simulation until the code crashed, that is a few times after the collision. The evolution of the contour surfaces of the lapse $\alpha$ are reported in the top row of figure \ref{headon}. The panels $(\hbox{a})$, $(\hbox{b})$ and $(\hbox{c})$ show the $x-y$ section of the 3D domain, at $z = L_0/2$, i.e. at the middle of the $z-$domain, at times $t=1,\, 13$ and $25 M$, respectively. As it can be seen, after the BH's collision, spurious harmonics develop and grow in time. These numerical disturbances are related to gravitational fluctuations, generated at the central collision-region, that propagate away and interact at the periodic borders. This non-physical interference is also amplified by the violation of the mean gravitational profile of the black holes at the border, which eventually leads the simulation to crash at $t \sim 30$.

In order to stabilize the spectral algorithm, a second simulation (RUN$_{35}$) has been carried out by switching-on the IHB technique. In order to quickly truncate noise that periodically travels through the periodic boundaries, we match the two solutions, $f^{\{ideal\}}({\bf x}, t)$ and $f^{\{H\}}({\bf x}, t)$, with a rigid condition
\begin{eqnarray}
	f({\bf x}, t)= \cases{f^{\{ideal\}}({\bf x}, t)&for $ |{\bf x}-{\bf x}_0|\leq \lambda $, \\
	f^{\{H\}}({\bf x}, t)&for $|{\bf x}-{\bf x}_0| > \lambda$,}
\nonumber
\end{eqnarray}
where ${\bf x}_0$ is set at the center of the 3D spatial domain, and $\lambda = \frac{3}{8}L_0$. In the bottom row of figure \ref{headon} we represent the contour surfaces of the lapse [panels $(\hbox{d}),(\hbox{e})$ and $(\hbox{f})$], at the same times of the previous run. The simulation has been noticeably improved, thanks to the diffusive boundaries that absorb the spurious periodic effects, as well as the out-propagating gravitational fluctuations. The line contours reveal a very smooth and well-behaved profile of the final (merged) black hole, confirming that dealiased pseudo-spectral methods can represent a good numerical strategy for the simulation of compact object dynamics.

Finally, we compare the violation of the Hamiltonian constraints for the two runs (with and without IHB), as reported in figure \ref{h2headon}. As it can be seen, the run with the viscous boundaries is much more stable, manifesting also lower violation of the constraints. This is due to the fact that the boundary ripples have been absorbed and the method prevents the contamination of these wiggles in the central ideal part of the domain, where the merger occurs. A similar strategy can be used for a variety of studies, including the inspiring binaries and the multiple black holes systems \cite{binaryBH, multipleBH, multipleBH2}.
\reply{ 
    There are other techniques able to handle black hole dynamics, as in the SpEC code \cite{ScheelEAL06, Szilagyi14}, where the method imposes a large infrastructure able to handle the singularity. The method presented here is different and simpler than the above (more refined) strategy. In the case of massive objects, we base our numerical approach on (1) a simple grid-shift of the black hole (in the case of the static Schwarzschild), (2) the use of filtering/dealiasing techniques that retain the differentiability of the solution and (3) the use of hyperviscous boundaries which absorb perturbations.
}

\section{Conclusions}
\label{concl}
\reply{
    Pioneering works on spectral methods applied to numerical relativity, as in the case of the SpEC and the SGRID codes, suggested that the spectral approach is very efficient. These codes, together with more common approaches such as BAM \cite{BrugmannEA04}, the Einstein Toolkit \cite{EToolkit1, EToolkit2}, and more exotic codes such as Dendro-GR \cite{FernandoEA2019}, represent today a valid tool of investigation for the understanding of gravitational dynamics. In this work we pursued the path of spectral algorithms, by examining the role of numerical instabilities and the performances of filtering methods.
}

We have studied the evolution of gravitational fields by solving numerically the Einstein field equations, in vacuum, with two different approaches, namely the ADM and the BSSN formalism. 
Our numerical method is based on a pseudo-spectral technique, where we compute spatial derivatives via simple and efficient Fast Fourier Transforms. A detailed analysis of the aliasing instabilities has been presented and two types of anti-aliasing filters have been used in order to suppress numerical artifacts that are due to the intrinsic nonlinear nature of the governing equations. We presented a new technique that improves the stability of the simulations, named the Running Stability Check, which monitors the strength of the time derivatives, providing an optimal time-step. The code shows a very low violation of the constraints, due to the precision of standard spectral methods, which are high-order accurate. We have found that the filter type and the cutoff wavenumber are crucial for the evolution of the gravitational equations, with very stable and accurate solutions, especially for smooth filters with cutoffs $k^*< N/3$. 

The stabilized, pseudo-spectral code can handle even more difficult gravitational problems, by solving BH equilibria and the dynamics of binary systems. In this regard, we did overcome the typical limitations of periodicity, which is needed for such pseudo-spectral, FTT-based methods. Our strategy, borrowed from fluid and plasma dynamics, is able to suppress spurious boundary effects. We matched the ideal solution in the center of the domain, based on the pseudo-spectral solution of the BSSN system of equations, with a semi-implicit, second order Crank-Nicholson technique at the boundaries, where we added hyper-viscous diffusion. This strategy has been successfully tested via a simple gravitational wavepacket absorption, as well as more catastrophic initial data, such as the head-on collision of black holes. 

In summary, the code has been successfully tested via the classical numerical relativity testbeds. Via these typical initial data, we have validated the goodness and the robustness of our numerical strategy. Future works will concentrate on the approach of the present algorithm to the dynamics of in-spiraling binary systems, as well as to the dynamics of many-body problems. Finally we plan to export the method to the solution of the gravitational fields in presence of matter as well its coupling with electromagnetic fields, in the framework of general relativistic magnetohydrodynamics.

\reply{

\appendix

\section{}

\subsection{The time integration scheme}

\begin{figure}[htbp]
\hspace{-4pt}
\includegraphics[height=95mm,width=148mm]{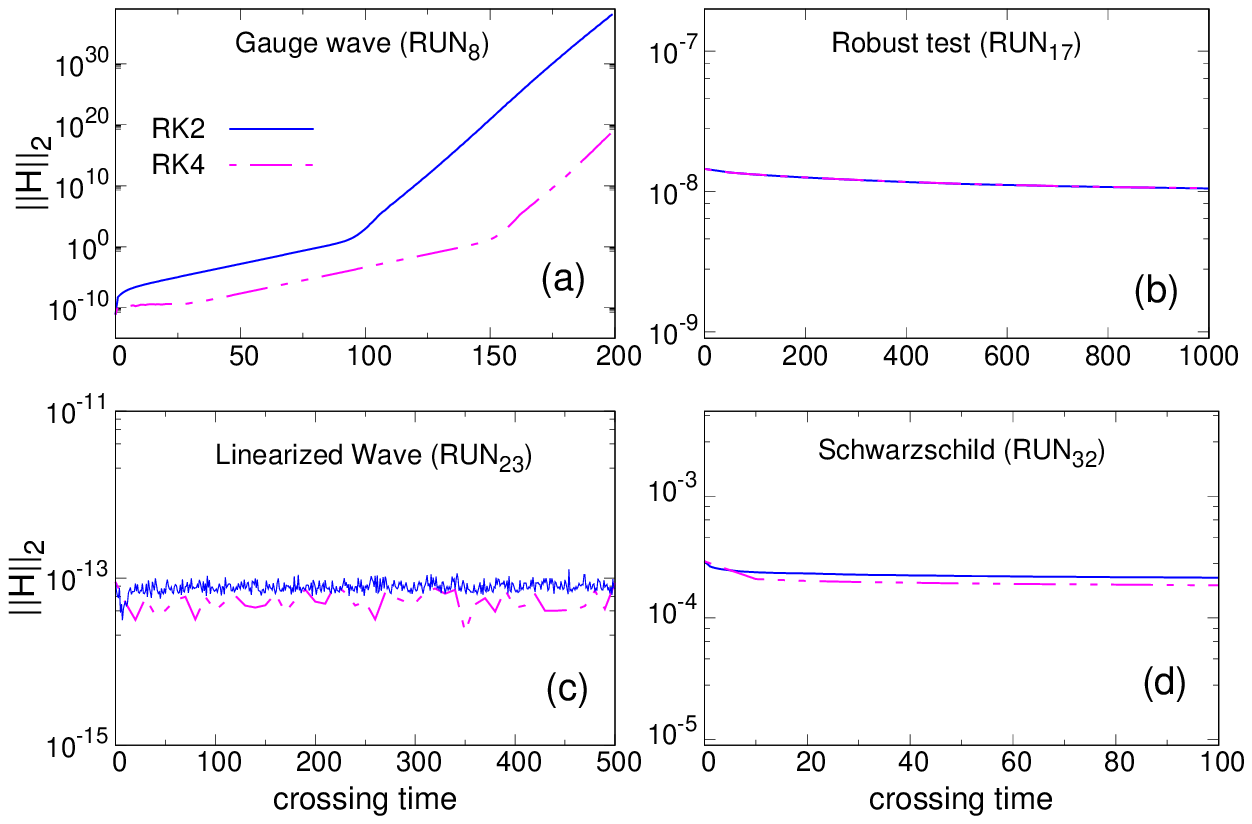}
\caption[ ]{\footnotesize{} Hamiltonian constraint for different initial data, for both second-order  (full) and fourth-order (dot-dot-full) Runge-Kutta. The performances of the runs are summarized on table \ref{table2}. 
}
\label{RK2RK4}
\end{figure}

The most commonly used methods to integrate in time systems of coupled partial differential equations are the well-known Runge-Kutta (RK) schemes. Several classifications of these methods can be done, according to their convergence order, the number of stages or their explicit/implicit structure. Low-order RK methods have the advantage of simplicity and the relatively low need of memory allocation. This is surely a good point for numerical relativity, especially in the BSSN formalism, where the number of  tensors proliferates compared to the ADM approach. However, low-order RK, such as the second-order method (RK2) are quite dissipative, causing a departure (in time) from exact solutions (when these are known).  Some of these accuracy problems are partially solved by using higher-order, such as the popular fourth-order RK4 \footnote{Alternatively, explicit (e.g. Adams-Bashford) or implicit (Adams-Moulton) linear multi-step methods improve the stability properties, albeit at the expense in increasing the complexity of the algorithm.}.

In this brief subsection of the Appendix we compare the accuracy of both RK2 and RK4, in the case of our pseudo-spectral filtered method. The RK2 has been described in Section \ref{numsec}, by equation (\ref{rk2}). Analogously, for the RK4 one gets
\begin{equation}
    f^{n+1}=f^{n}+\frac{\Delta t}{6}\left[ F_1 + 2 F_2 + 2 F_3 + F_4 \right], 
\label{rk4}
\end{equation}
where $F_1=F(f^n)$, $F_2=F(f^n+\frac{\Delta t}{2}F_1)$, $F_3=F(f^n+\frac{\Delta t}{2}F_2)$ and $F_4=F(f^n+ \Delta t F_3)$.

In figure \ref{RK2RK4} we show the Hamiltonian error $||H||_2$, for the most stable runs shown in table \ref{table}. For the sake of simplicity, we show only few testbeds, namely: the large amplitude Gauge wave (Run$_{8}$), robust stability test (Run$_{17}$), the gravitational wave test (Run$_{23}$) and the Schwarzschild black hole (Run$_{32}$). In all the cases, the RK4 has always a smaller violation error, as expected.


\begin{table}
\caption{\label{table2}Convergence table for different integration schemes. The numerical values in the table refers to the discrepancy $||\Delta_a \chi||_2$, defined in equation (\ref{eq:an}), evaluated for the second ($\#=2$) and the fourth ($\#=4$) order Runge-Kutta. The gauge waves have been compared at $t=100$, the gravitational waves at $t=500$ and the Schwarzschild BH at $t = 100$. }
\begin{indented}
\item[]\begin{tabular}{@{}llll}
\br
 Algorithm &  gauge wave  &   linearized wave &  Schwarzschild \\
\mr
  RK2   &   {$9.6\times 10^{-1}$}   &    {$2.4\times 10^{-12}$}   &    {$1.1\times 10^{-3}$} \\ 
  RK4   &   {$2.4\times 10^{-2}$}   &    {$2.3\times 10^{-12}$}   &    {$1.1\times 10^{-3}$} \\ 
\br
\end{tabular}
\end{indented}
\end{table}

%
%

In table \ref{table2} we report the difference between the simulated profile of the field $\chi$ and its analytical solution. In particular, we computed the $L_2$ norm of the error
\begin{equation}
\Delta_a \chi = \chi^{RK\#} - \chi^{analytical}, 
\label{eq:an}
\end{equation}
for both the second (\#=2) and the fourth (\#=4) order. 
    As it can be seen, the RK4 is more accurate, although it requires extra computational time and memory. In future works, a combination of high-order time-integration schemes with the IHB will be presented, coupling alternative techniques with implicit diffusion for the boundaries.



\subsection{Convergence tests}

In this second Appendix we present the convergence tests at different resolutions $N$, for the filtered BSSN algorithm. We start with the Schwarzschild black hole, discussed in subsection \ref{schws}, by varying the resolution from very-low ($N=32^3$) to moderate resolution ($N=256^3$). For this set of runs we used the same parameters as for RUN$_{32}$ (see table \ref{table}) and by shifting the puncture position in between mesh-nodes, as described before.

\begin{table}
\caption{\label{table3}Convergence table for the Schwarzschild black hole. The errors have been quantified as the $L_2$ norm of the discrepancy from the analytical solution,  similarly to equation (\ref{eq:an}). }
\begin{indented}
\item[]\begin{tabular}{@{}lllll}
\br
   Resolution &  $32^3$  &  $64^3$ &  $128^3$  &  $256^3$  \\
\mr
  $||\Delta_a \chi||_2$   &   {$1.5\times 10^{-3}$}   &    {$1.2\times 10^{-3}$}   &    {$1.1\times 10^{-3}$}  &    {$8.9\times 10^{-4}$} \\ 
\br
\end{tabular}
\end{indented}
\end{table}

\begin{figure}[htbp]
\hspace{-4pt}
\includegraphics[height=55mm,width=158mm]{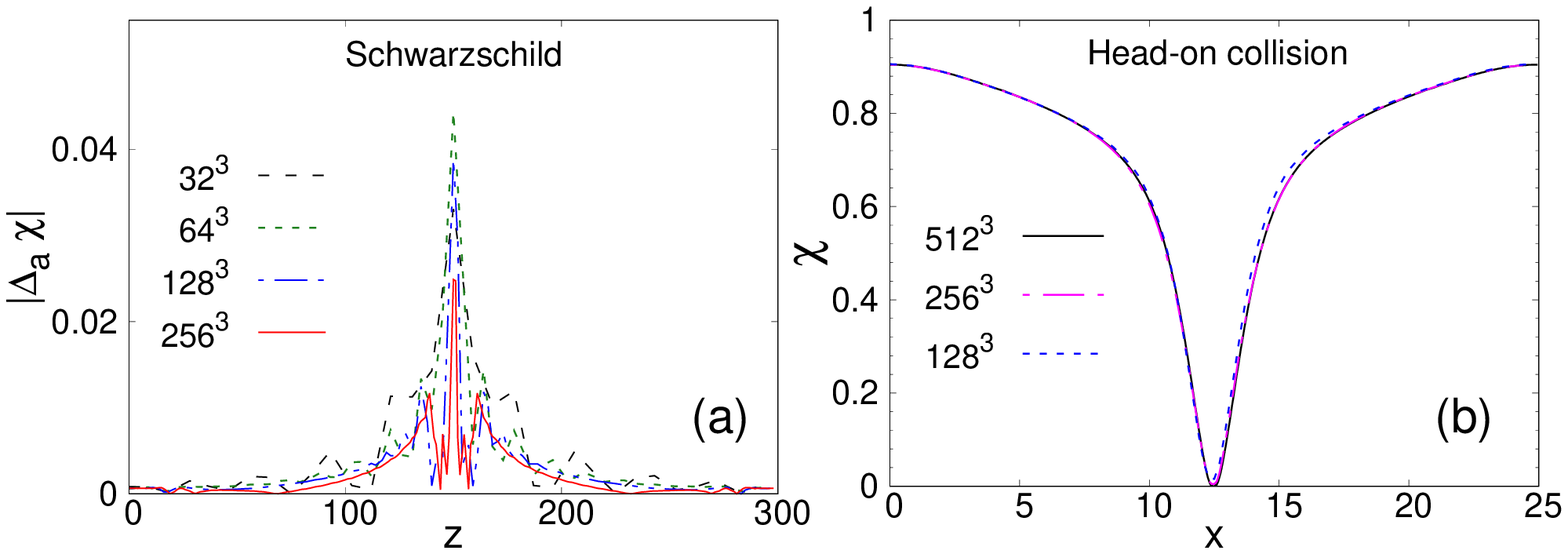}
\caption[ ]{\footnotesize{} (a) One dimensional cut of the discrepancy between the simulation and the analytical results $|\Delta_a \chi|$ (see text), for different resolutions summarized in table \ref{table3}. (b) Head-on collision, final profile, for different resolutions.}
\label{convrg}
\end{figure}

In order to quantify accuracy, for each resolution we computed the difference between a profile at the end of the simulation and its analytical solution, as in equation (\ref{eq:an}), where now $\Delta_a \chi = \chi^{N}-\chi^{analytical}$ and $\chi^{N}$ is the conformal factor for the simulation at a given resolution $N$. This norm can be computed for any variable but here we report results for $\chi$ (other fields give similar results, not shown). In figure \ref{convrg} (a) we show this discrepancy, for different numerical resolutions, at the end of the Schwarzschild simulations ($t=100$). As it can be seen (and as expected) the error diminishes with increasing resolution. To quantify the convergence, we computed the $L_2$ norm of this error, as reported in table \ref{table3}. The global error slightly diminishes, going from low to high resolution.

    As a final set of convergence tests, we repeated the head-on collision of black holes, discussed in the subsection \ref{eado}. In particular, we performed RUN$_{35}$, going from $128^3$ up to moderately-high resolution, with $N=512^3$ mesh points. We compared the simulations just after the collision of the BHs, at $t=22$, reported in panel (b) of figure \ref{convrg}. This sequence of runs  suggests that the solution is well described within our typical resolutions, for $N\geq 128^3$. In particular, the profiles are very similar, although the highest resolutions reveal a more singular and accurate profile. After the merging, the filter, together with the good differentiability properties of the BSSN formalism \cite{camp}, is able to handle the final singularity at the location of the puncture, which typically is expected to cause problems for spectral codes. To quantify the level of convergence here, we computed the error $\Delta_N \chi = \chi^{N_1} -\chi^{N_2}$, for neighboring spectral resolutions $N_1$ and $N_2$. In order to perform these direct measurements, we computed the difference only on the coarser grid and we used for all the simulations the same time-step, constrained by the $512^3$ run. We obtained $||\chi^{256} -\chi^{128}||_2 =3.3 \times 10^{-3}$ and  $||\chi^{512} -\chi^{256}||_2=9.7 \times 10^{-4}$. This further confirms that the solution is accurate and convergent.
}

\ack
The authors would like to thank Luciano Rezzolla for stimulating discussions and Sebastiano Bernuzzi for useful advice on the literature. The simulations have been performed 
\reply{
    by using a parallel architecture based on MPI directives, 
}
at the Newton HPPC Computing Facility at the University of Calabria. 


\end{document}